\documentclass[%
amsmath,amssymb,
aps, physrev,
]{revtex4-2}

\usepackage{xcolor}

\usepackage[T1]{fontenc}
\usepackage{array}
\usepackage{multirow}
\usepackage{amsmath}
\usepackage{slashed}
\usepackage{IEEEtrantools}
\usepackage{amstext,amssymb}
\usepackage{subcaption}
\usepackage{graphicx}
\usepackage{graphicx}
\usepackage{graphicx,wrapfig}
\usepackage{float}
\usepackage{orcidlink}
\usepackage{comment}
\usepackage{hyperref}
\hypersetup{
    colorlinks=true,
    citecolor=red,
    linkcolor=black,
    urlcolor=blue
}
\definecolor{mymaroon}{RGB}{120,45,45}

\begin{document}


\title{
\textcolor{mymaroon}{Neutrino Oscillation and Dark Matter Phenomenology in the Scotogenic Model with Non-Holomorphic Modular \(A_4\) Symmetry}
}

\author{Nilima Chakra}
\email{physicsnilima@gmail.com}
\affiliation{$^1$Department of Physics and Astronomical Science, Central University of Himachal Pradesh, Dharamshala (HP) 176215, India}

\author{Priya}
\email{kashyappriya963@gmail.com}
\affiliation{$^1$Department of Physics and Astronomical Science, Central University of Himachal Pradesh, Dharamshala (HP) 176215, India}

\author{B. C. Chauhan}
\email{bcawake@hpcu.ac.in}
\affiliation{$^1$Department of Physics and Astronomical Science, Central University of Himachal Pradesh, Dharamshala (HP) 176215, India}

\author{and Takaaki Nomura}
\email{nomura@scu.edu.cn}
\affiliation{$^2$College of Physics, Sichuan University, Chengdu 610065, China}
\begin{abstract}
Neutrino oscillation experiments have established that neutrinos are massive, yet the origin of their masses, the flavour structure of lepton mixing, and the nature of dark matter remain unresolved. Non-holomorphic modular symmetry, in which Yukawa couplings are polyharmonic Maa\ss{} forms controlled by a single complex modulus, offers a flavon-free way to fix the flavour structure, but its implications for the dark sector of radiative neutrino mass models have not been examined against the latest oscillation data. We study a scotogenic model with non-holomorphic modular $A_4$ symmetry in which an accidental $Z_2$ symmetry forbids tree-level neutrino masses and stabilises the dark-matter candidate, which can be either the lightest singlet Majorana fermion or the neutral inert scalar. A $\chi^2$ fit to the NuFIT~6.1 global data in normal ordering shows that the oscillation observables are reproduced only within a restricted region of the modulus, yielding predictions for the CP-violating phases while keeping the sum of neutrino masses and the effective Majorana mass below current cosmological and neutrinoless double-beta decay bounds. Because only a subset of the scalar-sector parameters enters the neutrino mass, the remaining ones can be adjusted to reproduce the observed relic density in both the fermionic and bosonic scenarios without altering the oscillation fit, while satisfying collider and charged-lepton flavour-violation constraints. These results show that non-holomorphic modular flavour symmetry can provide a predictive description of neutrino masses and mixing together with a viable dark-matter candidate using few free parameters, motivating further tests of such frameworks in future oscillation, neutrinoless double-beta decay, and dark-matter experiments.
\end{abstract}

\maketitle

\section{INTRODUCTION}

The discovery of neutrino oscillations has provided compelling evidence for physics beyond the Standard Model (SM) of particle physics. In its minimal formulation, the SM accommodates neutrinos as massless left-handed Weyl fermions. However, the discovery of neutrino flavor conversion by Super-Kamiokande~\cite{Super-Kamiokande:1998kpq}, the Sudbury Neutrino Observatory (SNO)~\cite{SNO:2002tuh}, and KamLAND~\cite{KamLAND:2002uet} established that neutrinos possess non-zero masses, revealing a fundamental limitation of the SM. Over the past two decades, measurements from solar, atmospheric, reactor, and accelerator-based neutrino experiments have significantly improved our understanding of the lepton sector, providing increasingly precise constraints on the three leptonic mixing angles, $\theta_{12}$, $\theta_{13}$, and $\theta_{23}$, and the solar and atmospheric mass-squared differences, $\Delta m^2_{\rm solar}$ and $\Delta m^2_{\rm atm}$. Nevertheless, the origin of neutrino masses, the mechanism underlying their generation, and the Dirac or Majorana nature of neutrinos remain unresolved. Addressing these questions is essential for understanding the possible structure of new physics beyond the SM and continues to motivate theoretical investigations of neutrino mass models.

Flavor symmetries have been widely investigated as a theoretical approach to understanding the observed structure of neutrino masses and lepton mixing ~\cite{Grossman:1998jj, Barman:2025bru, Chauhan:2023faf, Priya:2025khf, Priya:2026ytf, Tapender:2023kdk, Tapender:2025cfc, CarcamoHernandez:2026pis, deMedeirosVarzielas:2025byb, deMedeirosVarzielas:2026lfw, Ishiguro:2022pde, CarcamoHernandez:2025atu}. In conventional discrete flavor-symmetry frameworks, reproducing the observed fermion mass and mixing patterns often requires the introduction of multiple flavon fields, whose vacuum expectation values determine the structure of the fermion mass matrices. The implementation of suitable vacuum alignments, often supplemented by auxiliary symmetries to suppress unwanted interactions, can make the theoretical construction more involved and introduce additional model-dependent parameters~\cite{Chauhan:2023faf}.

Modular symmetry provides an alternative approach to constructing flavor models, where the fermion mass matrices are determined by the modular transformation properties of the fields and the associated modular forms. The complex modulus \(\tau\) acts as the central parameter governing the flavor structure. Unlike conventional flavor-symmetry models, this approach does not require the introduction of flavon fields to generate the desired flavor structure.
The framework is also motivated by string compactifications, in which modular symmetries can arise naturally from the underlying geometry~\cite{Feruglio:2017spp}. Models based on modular symmetry are extensively studied in the literature~\cite{Feruglio:2017spp, Ohki:2020bpo, deAdelhartToorop:2011re, Feruglio:2021dte, Kashav:2024lkr, Singh:2024imk, Kashav:2021zir, Kumar:2023moh, Mishra:2023ekx, Kobayashi:2021pav, Ishiguro:2022pde}.

The non-holomorphic modular framework extends the conventional modular symmetry approach by allowing modular forms to depend on both the modulus $\tau$ and its complex conjugate $\bar{\tau}$. This generalization is formulated in terms of polyharmonic Maaß forms, which allow Yukawa couplings to carry negative modular weights~\cite{Qu:2024rns}. The inclusion of negative-weight modular forms provides greater flexibility in constructing flavor models and offers additional possibilities for explaining fermion masses and mixing patterns. In recent years, non-holomorphic modular symmetries have been explored in several neutrino mass generation mechanisms, including Type-I~\cite{Cheshta:2026fls,Nanda:2025lem}, Type-II~\cite{Nomura:2024atp}, and Type-III seesaw scenarios~\cite{Priya:2025wdm, Priya:2026ehe}, as well as the scotogenic framework~\cite{Nomura:2024vzw}. Several other studies have further explored their applications in different aspects of neutrino phenomenology and flavor model building~\cite{Dey:2025zld,Kumar:2024uxn,Kumar:2025nut,Ding:2024inn,Li:2024svh,Loualidi:2025tgw,Li:2025kcr,Nomura:2024nwh,Okada:2025jjo,Abbas:2025nlv,Gao:2025jlw,Jangid:2025thp,Nomura:2025ovm,Nomura:2025raf,Zhang:2025dsa,Nasri:2026nbf,Tapender:2026ets,Zhang:2026kyy,Majhi:2026jdk,Loualidi:2026pld,Abbas:2026siw,Behera:2026tmo,Li:2026ory}.
Notice that the non-holomorphic modular framework does not require supersymmetry, which allows us to construct simpler models.  
In particular, the framework is suitable for realizing the structure of the original scotogenic model~\cite{Ma:2006km}. 


In this work, we extend the analysis of the scotogenic model with non-holomorphic $A_4$ symmetry, presented in Ref.~\cite{Nomura:2024vzw}, by incorporating updated data on neutrino phenomenology and exploring the implications of the model for the dark sector. In particular, we explore the relic abundance of a dark matter candidate considering both fermionic and bosonic cases in the model, which was not discussed in the previous research in detail. Furthermore, we perform a $\chi^2$ analysis using the latest neutrino oscillation data from NuFIT 6.1 to assess the compatibility of the model with current experimental constraints and obtain its preferred parameter space. Beyond neutrino phenomenology, we investigate the dark sector of the model, thereby broadening its scope and examining its implications for dark matter. This combined analysis provides a more comprehensive assessment of the model in light of current neutrino oscillation data and its potential to address the dark matter problem.

Combining non-holomorphic modular symmetry with the scotogenic mechanism is particularly interesting because it brings together two complementary aspects of neutrino phenomenology: the radiative origin of neutrino masses and the symmetry-based determination of their flavor structure. In a scotogenic framework, neutrino masses arise through loop diagrams involving new particles, while the modular symmetry can constrain the interactions and coupling matrices entering the loop-generated neutrino mass matrix. This combination therefore offers a systematic way to investigate whether the observed neutrino oscillation parameters can be accommodated within a more constrained theoretical framework. It also provides a setting in which the implications of the modular flavor structure for other phenomenological aspects of the model can be explored. The extent to which this construction improves predictivity depends on the field assignments, modular weights, and independent parameters retained in the model. 

The remainder of this paper is organised as follows. In Sec.~\ref{sec:model}, we present the scotogenic model with non-holomorphic modular $A_4$ symmetry, including the field content and modular weights, the modular-invariant Yukawa and Majorana mass terms, the charged-lepton sector, and the inert-scalar sector. We also give the one-loop neutrino mass formula, the observables for neutrinoless double-beta decay and cosmology, and the constraints from charged-lepton flavor violation and collider searches. Sec.~\ref{sec:results} contains the numerical results. In Subsec.~\ref{subsec:bosonic}, we discuss the bosonic dark-matter scenario, where $\eta_I$ is the lightest $Z_2$-odd particle, covering the neutrino oscillation fit and the relic-density analysis. In Subsec.~\ref{subsec:fermionic}, we do the same for the fermionic scenario, where $N_1$ is the lightest $Z_2$-odd particle. Sec.~\ref{sec:conclusion} summarizes our conclusions. Appendix~\ref{sec:A4conv} collects the $A_4$ multiplication rules and the $q$-expansions of the weight-zero polyharmonic Maa\ss{} form triplet, and Appendix~\ref{sec:tauequiv} discusses equivalent values of the modulus $\tau$.

\section{MODEL AND CONSTRAINTS}
\label{sec:model}

We consider a scotogenic model with non-holomorphic modular $A_4$ flavour symmetry~\cite{Nomura:2024vzw}. Besides the SM gauge group $SU(3)_C\times SU(2)_L\times U(1)_Y$, the theory has a finite modular $A_4$ symmetry ($\Gamma_3\simeq A_4$) under which each field carries an $A_4$ representation and a modular weight $k_I$. The SM lepton sector is extended by three gauge-singlet Majorana fermions $N_R$ and an inert scalar doublet $\eta$. The field content and charge assignments of the non-holomorphic modular $A_4$ scotogenic model are given below. 
\begin{table}[h]
\centering
\begin{tabular}{c|ccccc}
\hline\hline
 & $\overline{L_L}$ & $e_R$ & $N_R$ & $H$ & $\eta$ \\
\hline
$SU(3)_C$ & $1$ & $1$ & $1$ & $1$ & $1$ \\
$SU(2)_L$ & $2$ & $1$ & $1$ & $2$ & $2$ \\
$U(1)_Y$ & $\frac12$ & $-1$ & $0$ & $\frac12$ & $-\frac12$ \\
$A_4$ & $\mathbf{3}$ & $\mathbf{3}$ & $\mathbf{3}$ & $\mathbf{1}$ & $\mathbf{1}$ \\
$k_I$ & $-1$ & $1$ & $0$ & $0$ & $1$ \\
\hline\hline
\end{tabular}

\label{tab:fieldcontent}
\end{table}

The modular weights are chosen such that the undesired operators $\overline{N_R}L_LH$ and $H\eta$ are forbidden by the modular $A_4$ symmetry itself, so that no additional $Z_2$ symmetry has to be imposed by hand. The theory nevertheless possesses an accidental $Z_2$ symmetry, under which $N_R$ and $\eta$ are odd and all SM fields are even. This matter parity can be identified with $(-1)^{k+2s}$ or $(-1)^{k}$, where $s$ is the spin; the latter choice also forbids $\overline{N_R}HL$, while $N_R$ cannot couple to quarks because of $SU(3)_C$. The $Z_2$ forbids tree-level neutrino masses, which therefore arise only at loop level, and it renders the lightest $Z_2$-odd particle stable, providing a dark-matter candidate ~\cite{nomura2019modular}.

The flavour structure of the model is controlled by the modulus $\tau=x+iy$, which lives in the upper-half plane $\mathbb{H}=\{\tau\in\mathbb{C}\,|\,\mathrm{Im}\,\tau>0\}$, on which $\Gamma=SL(2,\mathbb{Z})$ acts as
\begin{equation}
\tau\;\to\;\gamma\tau=\frac{a\tau+b}{c\tau+d},\qquad ad-bc=1 ,
\end{equation}
and is generated by $S:\tau\to-1/\tau$ and $T:\tau\to\tau+1$ with $S^2=(ST)^3=\mathbb{I}$~\cite{Feruglio:2017spp}. Imposing $T^3=\mathbb{I}$ yields the finite group $\Gamma_3\simeq A_4$, with $S^2=T^3=(ST)^3=\mathbb{I}$. A matter field $\phi^{(I)}$ of weight $k_I$ transforms as
\begin{equation}
\phi^{(I)}\;\to\;(c\tau+d)^{-k_I}\,\rho^{(I)}(\gamma)\,\phi^{(I)},
\end{equation}
where $\rho^{(I)}$ is an irreducible representation of $A_4$. The couplings are built from the weight-zero polyharmonic Maa{\ss} form triplet
\begin{equation}
Y_3^{(0)}(\tau)=\big(Y^{(0)}_{3,1},\,Y^{(0)}_{3,2},\,Y^{(0)}_{3,3}\big)^T\equiv(y_1,y_2,y_3)^T ,
\end{equation}
which transforms as an $A_4$ triplet, contains both holomorphic and non-holomorphic ($y=\mathrm{Im}\,\tau$) pieces, and plays the role of the Yukawa couplings. Its explicit $q$-expansion and our $A_4$ conventions are collected in Appendix~\ref{sec:A4conv}. The Yukawa terms are modular invariant when the corresponding operators are $A_4$ singlets of total modular weight zero, while kinetic and quadratic terms are made invariant by writing them as $|\partial_\mu\phi^{(I)}|^2/(-i\tau+i\bar\tau)^{k_I}$ and $|\phi^{(I)}|^2/(-i\tau+i\bar\tau)^{k_I}$, using $-\tau+\bar\tau\to|c\tau+d|^{-2}(-\tau+\bar\tau)$. Once $\tau$ is fixed, the factor $(-i\tau+i\bar\tau)^{k_I}$ is absorbed into the field normalisation and the free parameters. Since the Yukawa couplings are then fixed by $\tau$, the flavour structure is determined by the single complex modulus together with a small number of coupling constants.
The modulus is, however, defined only up to modular equivalence. Since \(Y_3^{(0)}(\tau+1)=\rho(T)\,Y_3^{(0)}(\tau)\) and \(Y_3^{(0)}(\tau+3)=Y_3^{(0)}(\tau)\), and since the Lagrangian is modular invariant, values of \(\tau\) related by \(SL(2,\mathbb{Z})\) transformations describe the same theory up to field redefinitions. Furthermore, for real couplings, the transformation \(\tau\to-\bar{\tau}\) complex-conjugates the mass matrices and reverses the signs of the CP-violating phases. Therefore, the allowed values of \(\tau\) are understood modulo these equivalences, as discussed in Appendix~\ref{sec:tauequiv}.

With these ingredients, the modular invariant Yukawa and Majorana mass terms are

\begin{align}
\mathcal{L}_{Y}
&= M_0\,[\overline{N_R^c} N_R]_{\mathbf 1}
+ M_3\,Y_3^{(0)}[\overline{N_R^c} N_R]_{\mathbf 3_S}
\nonumber\\
&\quad+y_\ell\,[\overline{L_L} e_R]_{\mathbf 1}H
+y_{\ell3}\,Y_3^{(0)}[\overline{L_L} e_R]_{\mathbf 3_S}H
+y'_{\ell3}\,Y_3^{(0)}[\overline{L_L} e_R]_{\mathbf 3_A}H
\nonumber\\
&\quad+y_D\,[\overline{L_L} N_R]_{\mathbf 1}\,\eta
+y_{D3}\,Y_3^{(0)}[\overline{L_L} N_R]_{\mathbf 3_S}\,\eta
+y'_{D3}\,Y_3^{(0)}[\overline{L_L} N_R]_{\mathbf 3_A}\,\eta
+\text{h.c.},
\label{eq:yukawa}
\end{align}

where the contractions of the products with $Y_3^{(0)}$ to an $A_4$ singlet are implicit. Using the contraction rules of Appendix~\ref{sec:A4conv}, the first line of Eq.~\eqref{eq:yukawa} gives the mass matrix of the heavy Majorana fermions,
\begin{equation}
M_N=M_3
\begin{pmatrix}
\tilde M_0+2y_1 & -y_3 & -y_2\\
-y_3 & 2y_2 & \tilde M_0-y_1\\
-y_2 & \tilde M_0-y_1 & 2y_3
\end{pmatrix} with
\qquad \tilde M_0\equiv\frac{M_0}{M_3},
\label{eq:MN}
\end{equation}
which is diagonalised by a unitary matrix $U_N$ according to
\begin{equation}
D_N=U_N^TM_NU_N=\mathrm{diag}(D_{N1},D_{N2},D_{N3}).
\end{equation}
The last line of Eq.~\eqref{eq:yukawa} gives the Yukawa interaction among $L_L$, $N_R$ and $\eta$, which takes the form $(\overline{L}_L\,y_\eta N_R)\,\eta$, where both the symmetric and the antisymmetric triplet contractions contribute, through $y_{D3}$ and $y'_{D3}$ respectively, to
\begin{equation}
y_\eta=
\begin{pmatrix}
y_D+2y_{D3}y_1 & -(y_{D3}-y'_{D3})y_3 & -(y_{D3}+y'_{D3})y_2\\
-(y_{D3}+y'_{D3})y_3 & 2y_{D3}y_2 & y_D-(y_{D3}-y'_{D3})y_1\\
-(y_{D3}-y'_{D3})y_2 & y_D-(y_{D3}+y'_{D3})y_1 & 2y_{D3}y_3
\end{pmatrix}.
\label{eq:yeta}
\end{equation}
In the mass basis of $N_R$ this matrix becomes $\tilde y_\eta=y_\eta U_N$.

The charged-lepton sector follows from the second line of Eq.~\eqref{eq:yukawa}. After electroweak symmetry breaking, with $\langle H\rangle=v/\sqrt2$ and $v\simeq246$~GeV, it yields the charged-lepton mass matrix
\begin{equation}
m_\ell= \frac{v}{\sqrt{2}}
\begin{pmatrix}
y_\ell+2y_{\ell3}y_1 & -(y_{\ell3}-y'_{\ell3})y_3 & -(y_{\ell3}+y'_{\ell3})y_2\\
-(y_{\ell3}+y'_{\ell3})y_3 & 2y_{\ell3}y_2 & y_\ell-(y_{\ell3}-y'_{\ell3})y_1\\
-(y_{\ell3}-y'_{\ell3})y_2 & y_\ell-(y_{\ell3}+y'_{\ell3})y_1 & 2y_{\ell3}y_3
\end{pmatrix},
\label{eq:ml}
\end{equation}
which is diagonalised by a bi-unitary transformation,
\begin{equation}
\mathrm{diag}(m_e,m_\mu,m_\tau)=V_L^\dagger m_\ell V_R,\qquad
V_L^\dagger\, m_\ell m_\ell^\dagger\, V_L=\mathrm{diag}(|m_e|^2,|m_\mu|^2,|m_\tau|^2).
\end{equation}
For a given $\tau$, the couplings $y_\ell$, $y_{\ell3}$ and $y'_{\ell3}$ are fixed by reproducing the observed charged-lepton masses through the three relations
\begin{align}
\mathrm{Tr}[m_\ell^\dagger m_\ell]&=|m_e|^2+|m_\mu|^2+|m_\tau|^2,\\
\mathrm{Det}[m_\ell^\dagger m_\ell]&=|m_e|^2|m_\mu|^2|m_\tau|^2,\\
\big(\mathrm{Tr}[m_\ell^\dagger m_\ell]\big)^2-\mathrm{Tr}\big[(m_\ell^\dagger m_\ell)^2\big]&=2\big(|m_e|^2|m_\mu|^2+|m_\mu|^2|m_\tau|^2+|m_e|^2|m_\tau|^2\big).
\end{align}

We now turn to the scalar sector, which controls both the radiative neutrino mass and the bosonic dark-matter scenario. The inert doublet $\eta$ acquires no vacuum expectation value, and after electroweak symmetry breaking the doublets take the form
\begin{equation}
H=\begin{pmatrix}0\\ \dfrac{v+h}{\sqrt2}\end{pmatrix},\qquad
\eta=\begin{pmatrix}\dfrac{1}{\sqrt2}(\eta_R+i\eta_I)\\[2mm] \eta^-\end{pmatrix}.
\end{equation}
The terms of the scalar potential relevant for the inert-sector spectrum are as follows.
\begin{equation}
V\supset\mu_\eta^2\,\eta^\dagger\eta+\lambda_{H\eta}(H^\dagger H)(\eta^\dagger\eta)+\lambda'_{H\eta}|H\eta|^2
+\big[\lambda\,(H\eta)^2+\text{h.c.}\big],
\qquad \lambda\equiv\lambda_0 \,\tilde Y_1^{(-2)},
\label{eq:V}
\end{equation}
where $\tilde Y_r^{(k)}$ denotes a modular form in the $A_4$ representation $r$ of weight $k$; here $\tilde Y_1^{(-2)}$ is a weight $-2$ singlet compensating the weight of $(H\eta)^2$. Further ``trivial'' terms are omitted. Taking $\lambda$ real, the physical masses are
\begin{align}
m_{\eta_R}^2&=\mu_\eta^2+\tfrac{v^2}{2}\big(\lambda_{H\eta}+\lambda'_{H\eta}+2\lambda\big),\\
m_{\eta_I}^2&=\mu_\eta^2+\tfrac{v^2}{2}\big(\lambda_{H\eta}+\lambda'_{H\eta}-2\lambda\big),\\
m_{\eta^\pm}^2&=\mu_\eta^2+\tfrac{v^2}{2}\lambda_{H\eta},
\end{align}
so that
\begin{equation}
\Delta m_\eta^2\equiv m_{\eta_R}^2-m_{\eta_I}^2=2\lambda v^2 .
\end{equation}
This splitting is essential for the radiative neutrino mass: for $\lambda\to0$ the neutral states are degenerate and $M_\nu$ vanishes. The neutrino sector depends on the four parameters $(\mu_\eta,\lambda_{H\eta},\lambda'_{H\eta},\lambda)$ only through the inert-scalar masses $m_{\eta_R}$ and $m_{\eta_I}$, entering the loop function in Eq.~\eqref{eq:Mnu}, and the charged-scalar mass $m_{\eta^\pm}$, entering the lepton-flavour-violation rates. It is therefore convenient to trade the four couplings for the physical set $\{m_{\eta_I},\,\Delta m_\eta^2,\,\delta m,\,\lambda_{H\eta}\}$, where we define the splitting $\delta m\equiv m_{\eta^\pm}-m_{\eta_I}$, so that $m_{\eta^\pm}=m_{\eta_I}+\delta m$ and $m_{\eta_R}^2=m_{\eta_I}^2+\Delta m_\eta^2$. Inverting the relations above gives
\begin{equation}
\lambda=\frac{\Delta m_\eta^2}{2v^2},\qquad
\lambda'_{H\eta}=\frac{\Delta m_\eta^2-2\,\delta m\,(2m_{\eta_I}+\delta m)}{v^2},\qquad
\mu_\eta^2=m_{\eta^\pm}^2-\frac{v^2}{2}\lambda_{H\eta},
\end{equation}
where we used $m_{\eta^\pm}^2-m_{\eta_I}^2=\delta m\,(2m_{\eta_I}+\delta m)$. In our neutrino scans we vary $m_{\eta_I}$ and $\Delta m_\eta^2$, which control the size of the radiative neutrino masses and hence the oscillation observables. The remaining two parameters, $\delta m$ and $\lambda_{H\eta}$, do not enter $M_\nu$ once $m_{\eta_I}$ and $\Delta m_\eta^2$ are given, and they are fixed only in the dark-matter analysis. Taking $\lambda>0$ corresponds to $\Delta m_\eta^2>0$, and the requirement that $\eta^\pm$ is heavier than $\eta_I$ corresponds to $\delta m>0$.

Neutrino masses then arise at one loop through the exchange of $N_R$ and $\eta_{R,I}$. In the mass basis of $N_R$ the mass matrix takes the standard scotogenic form~\cite{Ma:2006km},
\begin{equation}
(M_\nu)_{\alpha\beta}\simeq\sum_{k=1}^{3}\frac{(\tilde y_\eta)_{\alpha k}\,D_{Nk}\,(\tilde y_\eta)_{\beta k}}{(4\pi)^2}
\left[\frac{m_{\eta_R}^2}{m_{\eta_R}^2-D_{Nk}^2}\ln\frac{m_{\eta_R}^2}{D_{Nk}^2}
-\frac{m_{\eta_I}^2}{m_{\eta_I}^2-D_{Nk}^2}\ln\frac{m_{\eta_I}^2}{D_{Nk}^2}\right],
\label{eq:Mnu}
\end{equation}
with $\tilde y_\eta=y_\eta U_N$. The light-neutrino mass matrix is diagonalised as $V_\nu^TM_\nu V_\nu=D_\nu=\mathrm{diag}(m_1,m_2,m_3)$, and the PMNS matrix is
\begin{equation}
U_{\rm PMNS}=V_L^\dagger V_\nu .
\end{equation}
In the standard parametrisation, the mixing angles are obtained from
\begin{equation}
\sin^2\theta_{13}=|U_{13}|^2,\quad
\sin^2\theta_{12}=\frac{|U_{12}|^2}{1-|U_{13}|^2},\quad
\sin^2\theta_{23}=\frac{|U_{23}|^2}{1-|U_{13}|^2},
\end{equation}
and the CP-violating phases follow from
\begin{align}
J_{CP}&=\mathrm{Im}\big[U_{11}U_{22}U_{12}^\ast U_{21}^\ast\big]=s_{23}c_{23}s_{12}c_{12}s_{13}c_{13}^2\sin\delta_{CP},\\
\mathrm{Im}\big[U_{11}^\ast U_{12}\big]&=c_{12}s_{12}c_{13}^2\sin\tfrac{\alpha_{21}}{2},\qquad
\mathrm{Im}\big[U_{11}^\ast U_{13}\big]=c_{12}s_{13}c_{13}\sin\!\big(\tfrac{\alpha_{31}}{2}-\delta_{CP}\big).
\end{align}
The observables entering the neutrinoless double-beta decay and cosmological bounds are
\begin{equation}
m_{ee}=\Big|m_1c_{12}^2c_{13}^2+m_2s_{12}^2c_{13}^2e^{i\alpha_{21}}+m_3s_{13}^2e^{i(\alpha_{31}-2\delta_{CP})}\Big|,
\qquad \sum_i m_i=m_1+m_2+m_3 .
\end{equation}

The same spectrum determines the dark-matter phenomenology. The accidental $Z_2$ symmetry guarantees the stability of the lightest $Z_2$-odd particle, and the $Z_2$-odd spectrum consists of the three Majorana fermions $N_{1,2,3}$ with masses $D_{Nk}$ and the inert scalars $\eta_R$, $\eta_I$ and $\eta^\pm$, so that two distinct dark-matter scenarios arise. If $N_1$ is the lightest of these states, $D_{N1}<m_{\eta_R},m_{\eta_I},m_{\eta^\pm}$, dark matter is fermionic and communicates with the SM through the Yukawa matrix $\tilde y_\eta$, whose flavour structure is fixed by the modulus $\tau$. If instead the lightest $Z_2$-odd state is a neutral inert scalar, dark matter is bosonic. We take $\lambda>0$, so that $m_{\eta_R}>m_{\eta_I}$ and the dark-matter candidate is $\eta_I$, which must satisfy $m_{\eta_I}<D_{Nk},m_{\eta^\pm}$; it interacts through the gauge couplings, through the Higgs-portal couplings $\lambda_{H\eta}$, $\lambda'_{H\eta}$ and $\lambda$, and through $\tilde y_\eta$. The remaining free parameters $\delta m$ and $\lambda_{H\eta}$ are fixed in the dark-matter analysis.

We confront the model with a set of phenomenological constraints. The sum of the light neutrino masses must satisfy the upper bound
from cosmological observations. We consider both the constraint
from CMB data alone \cite{aghanim2020planck} and the stronger
constraint obtained when BAO data are included
\cite{Planck:2018vyg, aghanim2020planck,adame2025desi}. Since the Yukawa matrix $y_\eta$ couples the charged leptons to $N_R$ and $\eta^\pm$, the model also induces charged-lepton flavour violating decays. In the mass basis, with $Y'_\eta\equiv V_L^\dagger y_\eta U_N$, the branching ratios of $\ell_i\to\ell_j\gamma$ are approximately given by~\cite{toma2014lepton}
\begin{equation}
\mathrm{BR}(\ell_i\to\ell_j\gamma)\approx\frac{48\pi^3\alpha_{\rm em}C_{ij}}{G_F^2(4\pi)^4}\Big|\sum_{\alpha=1}^{3}Y'_{\eta\,j\alpha}Y'^{\dagger}_{\eta\,\alpha i}F(D_{N\alpha},m_{\eta^\pm})\Big|^2,
\end{equation}
with the loop function
\begin{equation}
F(m_a,m_b)\approx\frac{2m_a^6+3m_a^4m_b^2-6m_a^2m_b^4+m_b^6+12m_a^4m_b^2\ln(m_b/m_a)}{12(m_a^2-m_b^2)^4},
\end{equation}
where $C_{21}=1$, $C_{31}=0.1784$, $C_{32}=0.1736$, $\alpha_{\rm em}(m_Z)=1/128.9$ and $G_F=1.166\times10^{-5}$~GeV$^{-2}$. These branching ratios are required to lie below the current experimental upper limits. The charged inert scalar can be produced through electroweak processes at colliders, and we therefore require $m_{\eta^\pm}\gtrsim100$~GeV, following the LEP lower bound on charged scalars. Finally, the predicted $m_{ee}$ is compared with the current limit from KamLAND-Zen~\cite{KamLAND-Zen:2024eml} and with the projected sensitivities of future neutrinoless double-beta decay experiments such as nEXO~\cite{nEXO:2021ujk}.

In summary, the model is specified by the modulus $\tau$, the charged-lepton couplings $(y_\ell,y_{\ell3},y'_{\ell3})$, which are fixed by the observed charged-lepton masses, the fermion mass parameters $\tilde M_0=M_0/M_3$ and $M_3$, both of which are scanned, the Yukawa couplings $(y_D,y_{D3},y'_{D3})$, and the scalar parameters $(m_{\eta_I},\Delta m_\eta^2,\delta m,\lambda_{H\eta})$, of which the first two are scanned in the neutrino analysis, while $\delta m$ and $\lambda_{H\eta}$ are fixed in the dark-matter calculation.

\section{RESULTS\label{sec:results}}

We now present the numerical results for the two dark-matter scenarios allowed by the model, which differ in the identity of the lightest  $Z_2$-odd particle. In both scenarios the neutrino oscillation data are taken in normal ordering, and the points retained in the analysis satisfy the LEP bound on $m_{\eta^\pm}$ and the lepton-flavour-violation limits on $\ell_i\to\ell_j\gamma$. For each scenario we first show that the model reproduces the neutrino oscillation data and satisfies the neutrinoless double-beta decay and cosmological bounds. We then compute the relic density at fixed benchmark values of $\delta m$ and $\lambda_{H\eta}$, and finally show how these two parameters, which do not enter the neutrino sector, can be tuned to bring the points to the observed relic density.

The modulus is varied within $-1.5\le\mathrm{Re}\,\tau\le1.5$, which is one full period of $Y_3^{(0)}$ in $\mathrm{Re}\,\tau$, and $0<\mathrm{Im}\,\tau\le3$, and the remaining parameters are scanned logarithmically over $0.1\le\tilde M_0\le10$, $0.1\le M_3/\mathrm{GeV}\le10^{3}$, $10^{2}\le m_{\eta_I}/\mathrm{GeV}\le10^{5}$, $10^{-5}\le\Delta m_\eta^2/\mathrm{GeV}^2\le10$ and $10^{-8}\le y_D,\,y_{D3},\,y'_{D3}\le10$. The lower limit on $m_{\eta_I}$ ensures $m_{\eta_I}>100$~GeV, so that $m_{\eta^\pm}=m_{\eta_I}+\delta m>100$~GeV for $\delta m>0$ and the LEP bound is satisfied automatically in both scenarios.
Note also that we choose all the parameters except for $\tau$ to be real. This choice can be justified by assuming generalized CP symmetry in the non-holomorphic modular symmetry framework~\cite{Qu:2024rns}.

\subsection{Bosonic dark matter\label{subsec:bosonic}}
\begin{figure*}[tbp]
    \centering
    \includegraphics[width=\textwidth]{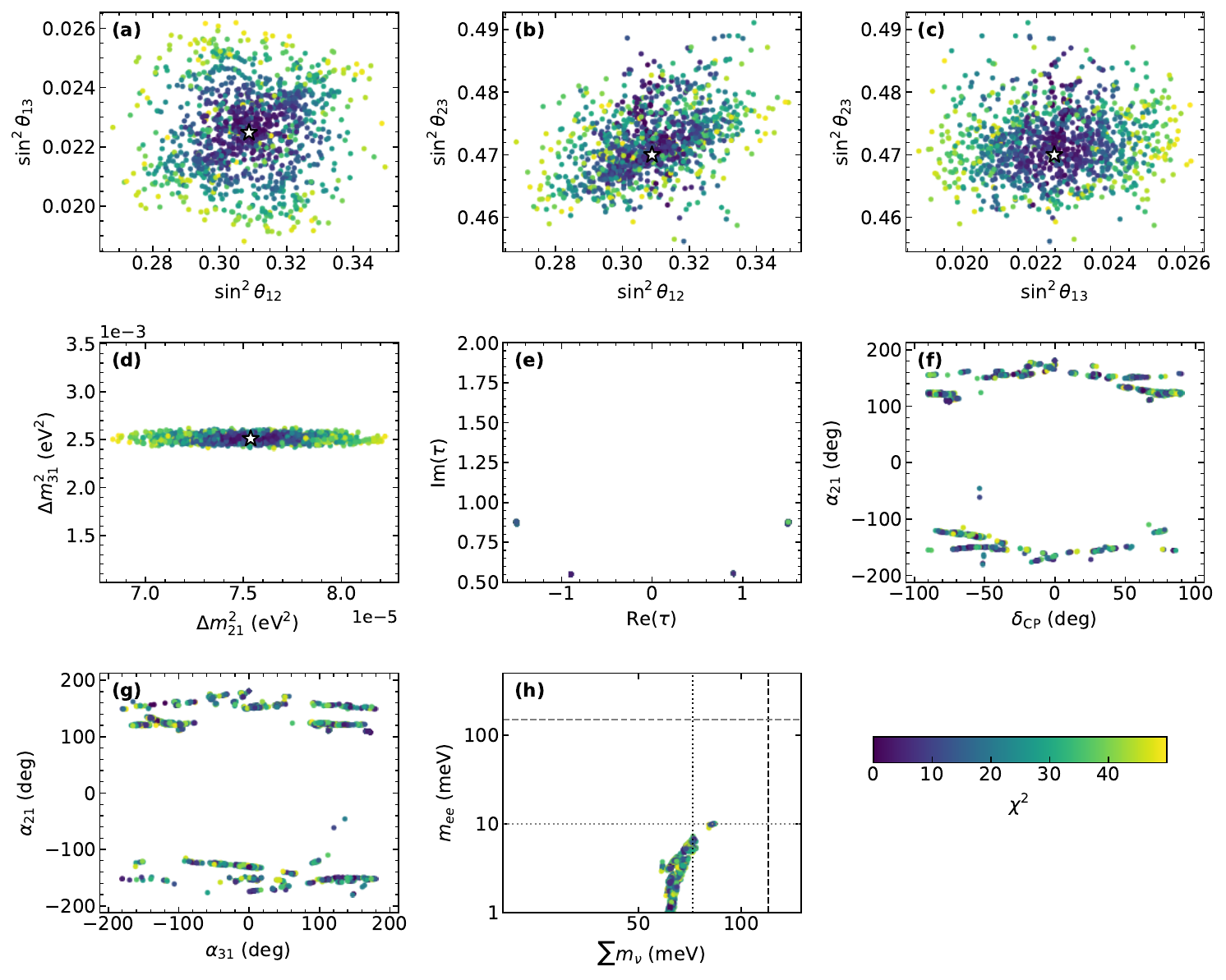}
    \caption{Neutrino oscillation phenomenology for the bosonic dark-matter scenario ($m_{\eta_I}<D_{N1}$), in normal ordering. Panels (a)--(c) show the correlations among $\sin^2\theta_{12}$, $\sin^2\theta_{13}$ and $\sin^2\theta_{23}$, and panel (d) the correlation between $\Delta m^2_{21}$ and $\Delta m^2_{31}$. Panel (e) shows the region of the complex modulus $\tau$ compatible with the oscillation data, panel (f) the correlation between $\delta_{CP}$ and $\alpha_{21}$, and panel (g) that between $\alpha_{31}$ and $\alpha_{21}$. Panel (h) shows $\sum_i m_i$ against $m_{ee}$, together with the KamLAND-Zen limit, the projected nEXO sensitivity, and the cosmological bounds from CMB and CMB+BAO data. The agreement with the data is quantified by the $\chi^2$ computed with $1\sigma$ uncertainties. Most of the points shown satisfy all the bounds, while some exceed the CMB+BAO bound in panel (h).}
    \label{fig:neutrino_oscillation}
\end{figure*}

We first consider the scenario in which $\eta_I$ is the lightest  $Z_2$-odd particle. All scans in this subsection therefore assume $m_{\eta_I}<D_{N1}$ and $\lambda>0$, with the parameters scanned over the ranges given above.
The agreement of each point with the oscillation data, taken in normal ordering, is quantified by a $\chi^2$ function built from the three mixing angles $\sin^2\theta_{12}$, $\sin^2\theta_{13}$ and $\sin^2\theta_{23}$ and the two mass-squared differences $\Delta m^2_{21}$ and $\Delta m^2_{31}$, using the NuFIT~v6.1 global fit~\cite{nufit61} with the uncertainties taken at the $1\sigma$ level.

The resulting neutrino phenomenology is summarised in Fig.~\ref{fig:neutrino_oscillation}. Panels (a), (b) and (c) show the correlations among $\sin^2\theta_{12}$, $\sin^2\theta_{13}$ and $\sin^2\theta_{23}$, and panel (d) the correlation between the two mass-squared differences $\Delta m^2_{21}$ and $\Delta m^2_{31}$. The points that satisfy the $\chi^2$ criterion populate the experimentally allowed region of each plane. Panel (e) shows that the accepted points cluster at a few isolated values of $\tau$, which, since the Yukawa couplings are fixed by $\tau$, is the origin of the predictive power of the model in the lepton sector. These clusters are not all physically distinct, because $\tau$ is defined only up to the equivalences discussed in Appendix~\ref{sec:tauequiv}. The clusters at $\mathrm{Re}\,\tau\simeq\pm1.5$ with $\mathrm{Im}\,\tau\simeq0.87$ lie on the same line, since $Y_3^{(0)}$ has period 3 in $\mathrm{Re}\,\tau$, and correspond to the neighbourhood of $\omega=e^{2\pi i/3}$, a fixed point of $ST$ at which a residual $Z_3$ symmetry is preserved. The clusters at $\mathrm{Re}\,\tau\simeq\pm0.9$ and $\mathrm{Im}\,\tau\simeq0.56$ are related to each other by $\tau\to-\bar\tau$ and are mapped by $\tau\to-1/(\tau\pm1)$ to $\mathrm{Im}\,\tau\simeq1.7$, in the region of large $\mathrm{Im}\,\tau$ where $Y_3^{(0)}$ is dominated by its first component. The accepted points therefore probe two physically distinct regions of the modulus, the neighbourhood of $\omega$ and the region of large $\mathrm{Im}\,\tau$.
 The results are basically consistent with the previous analysis in ref.~\cite{Nomura:2024vzw}, with some differences arising from the updated neutrino data and our choice of real parameters; in the previous analysis some parameters are taken to be complex. Thus our case is more predictive since the number of free parameters is reduced by eliminating complex phases.
The CP-violating phases do not enter the $\chi^2$, so panels (f) and (g) show the predictions of the model for $\delta_{CP}$ and the Majorana phases, through the correlations between $\delta_{CP}$ and $\alpha_{21}$ and between $\alpha_{31}$ and $\alpha_{21}$. Finally, panel (h) shows the sum of the neutrino masses against the effective Majorana mass $m_{ee}$, together with the KamLAND-Zen limit~\cite{KamLAND-Zen:2024eml}, the projected nEXO sensitivity~\cite{nEXO:2021ujk}, and the cosmological upper bounds on $\sum_i m_i$ from CMB data alone and from CMB combined with BAO data~\cite{Planck:2018vyg,aghanim2020planck,adame2025desi}. Most of the points that reproduce the oscillation data lie below each of these benchmark lines, so the model is consistent with the current neutrinoless double-beta decay and cosmological constraints; some points are in tension with the CMB+BAO limit.

\begin{figure*}[tbp]
    \centering
    \includegraphics[width=0.8\textwidth]{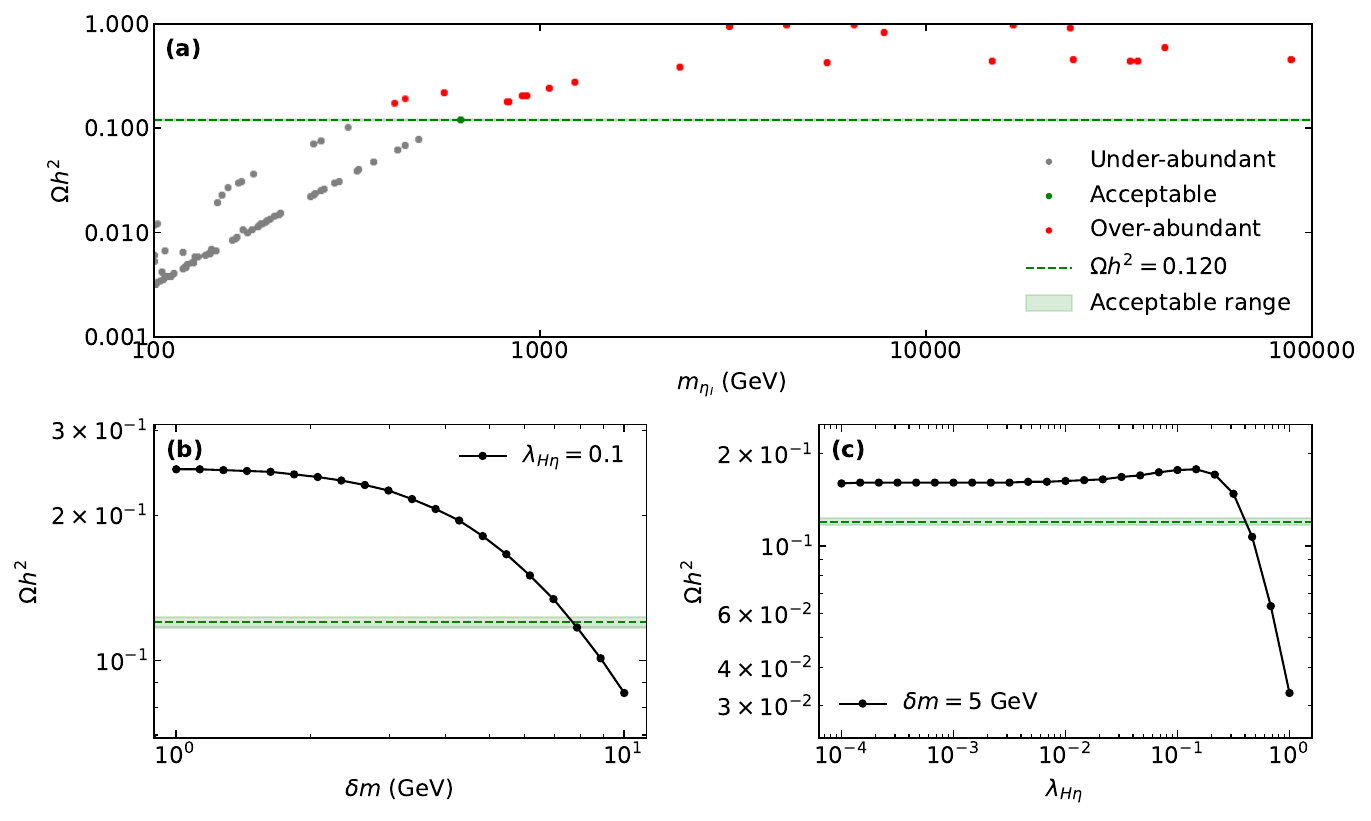}
    \caption{Relic density of $\eta_I$ for the bosonic dark-matter scenario. Panel (a) shows $\Omega h^2$ as a function of $m_{\eta_I}$ for the points that reproduce the neutrino oscillation data, with $\delta m=5$~GeV and $\lambda_{H\eta}=0.1$ held fixed. Panels (b) and (c) show $\Omega h^2$ for a representative point of panel (a), with one parameter varied while the other is kept at its benchmark value: (b) as a function of $\delta m=m_{\eta^\pm}-m_{\eta_I}$ at $\lambda_{H\eta}=0.1$, and (c) as a function of $\lambda_{H\eta}$ at $\delta m=5$~GeV. In all panels the green dashed line marks the observed value $\Omega h^2=0.120$ and the green band, labelled ``Acceptable'', the range $0.117\le\Omega h^2\le0.123$. In panel (a), underabundant points are shown in gray and overabundant points in red. The neutrino oscillation observables are unchanged along the curves in panels (b) and (c).}
    \label{fig:dm_BDM}
\end{figure*}

We now turn to the dark-matter phenomenology. The relic density of $\eta_I$ is determined by its annihilation through the gauge interactions and the Higgs-portal couplings, and by its coannihilation with $\eta_R$ and $\eta^\pm$. For every point of the neutrino scan, $m_{\eta_I}$ and $\Delta m_\eta^2$ are fixed by the oscillation fit, so that the mass of $\eta_R$ is already determined, and the only remaining freedom resides in $\delta m$ and $\lambda_{H\eta}$. After fitting the neutrino oscillation data, we implement the model using \texttt{FeynRules} (version 2.3.49)~\cite{alloul2014feynrules}, export it in the \texttt{CalcHEP} format, and compute the relic density using \texttt{micrOMEGAs} (version 7.1.4)~\cite{belanger2026micromegas}. Fixing $\delta m$ and $\lambda_{H\eta}$ to the benchmark values $\delta m=5$~GeV and $\lambda_{H\eta}=0.1$, we obtain the relic density as a function of $m_{\eta_I}$ shown in panel (a) of Fig.~\ref{fig:dm_BDM}. We take the observed value $\Omega h^2=0.120$ and regard a point as acceptable if $0.117\le\Omega h^2\le0.123$. At these benchmark values, most points lie outside the acceptable range, being either overabundant, $\Omega h^2>0.123$, or underabundant, $\Omega h^2<0.117$. A small fraction of points already satisfies the observed relic density at the benchmark values and requires no tuning, while all the others have to be tuned.
Examining the channel decomposition given by \texttt{micrOMEGAs} for several points, we find that the annihilation of $\eta_I\eta_I$ and $\eta_R\eta_R$ into $W^+W^-$ and $ZZ$ accounts for about half of $1/\Omega h^2$, and that most of the remainder comes from coannihilation channels involving $\eta^\pm$. These leading channels are largely unchanged for the tuned points.

The points outside the acceptable range can be brought to the observed relic density without altering the neutrino fit. The oscillation observables depend on the scalar sector only through $m_{\eta_I}$ and $\Delta m_\eta^2$, and hence they are independent of both $\delta m$ and $\lambda_{H\eta}$, so varying these two parameters leaves every point of Fig.~\ref{fig:neutrino_oscillation} unchanged.
The splitting $\delta m=m_{\eta^\pm}-m_{\eta_I}$ affects the relic density in two ways. It controls the coannihilation of $\eta_I$ with the charged scalar, which becomes less efficient as $\delta m$ grows, and, through the inversion relations of Sec.~\ref{sec:model}, it sets the Higgs-portal coupling of $\eta_I$, $\lambda_{\eta_I}\equiv(\lambda_{H\eta}+\lambda'_{H\eta}-2\lambda)/2=\lambda_{H\eta}/2-\delta m\,(2m_{\eta_I}+\delta m)/v^2$, whose magnitude grows rapidly with $\delta m$ for the masses considered here. Over the range of $\delta m$ explored, the second effect dominates, so that the relic density decreases as $\delta m$ increases. Overabundant points are therefore brought into the acceptable range by increasing $\delta m$, and underabundant points by decreasing it.
This is illustrated in panels (b) and (c) of Fig.~\ref{fig:dm_BDM} for a representative point, where one parameter is varied while the other is kept at its benchmark value, $\lambda_{H\eta}=0.1$ in panel (b) and $\delta m=5$~GeV in panel (c). For most points, varying $\delta m$ alone is sufficient to reach the acceptable range, whereas for a sizeable number of points the two parameters have to be tuned together, with $\lambda_{H\eta}$ supplying the additional change in the relic density that $\delta m$ cannot provide by itself. Since $\delta m$ changes $m_{\eta^\pm}$, we evaluate the branching ratios of $\ell_i\to\ell_j\gamma$ at the tuned values, and the points retained after tuning all satisfy the current experimental upper limits, so the tuning into the acceptable range does not conflict with the lepton-flavour-violation constraints.
\subsection{Fermionic dark matter\label{subsec:fermionic}}
\begin{figure*}[t]
    \centering
    \includegraphics[width=\textwidth]{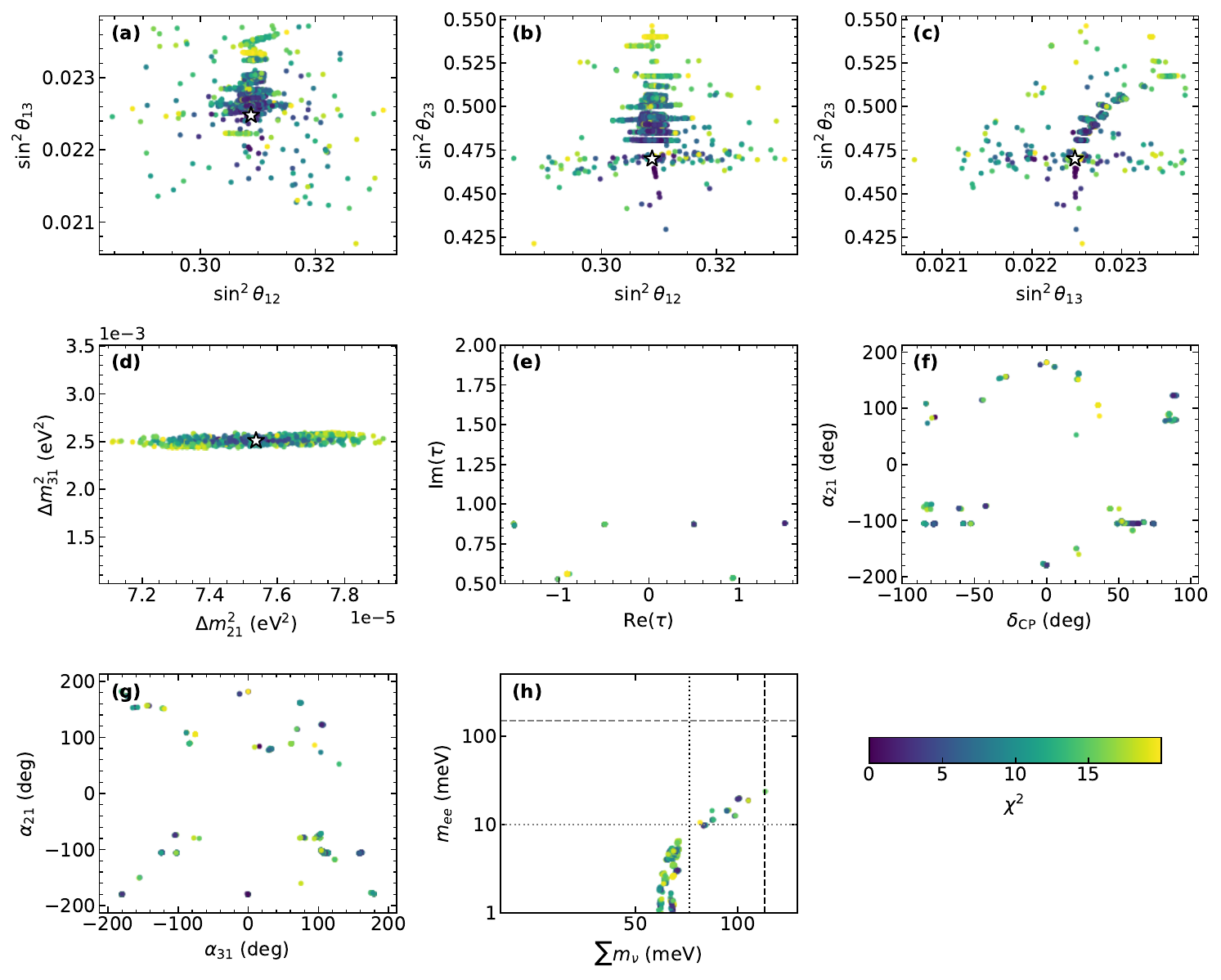}
    \caption{Neutrino oscillation phenomenology for the fermionic dark-matter scenario ($D_{N1}<m_{\eta_I}$), in normal ordering. The panels are as in Fig.~\ref{fig:neutrino_oscillation}: (a)--(c) the correlations among $\sin^2\theta_{12}$, $\sin^2\theta_{13}$ and $\sin^2\theta_{23}$, (d) the correlation between $\Delta m^2_{21}$ and $\Delta m^2_{31}$, (e) the region of the complex modulus $\tau$ compatible with the oscillation data, (f) the correlation between $\delta_{CP}$ and $\alpha_{21}$, (g) that between $\alpha_{31}$ and $\alpha_{21}$, and (h) $\sum_i m_i$ against $m_{ee}$ with the KamLAND-Zen limit, the projected nEXO sensitivity, and the CMB and CMB+BAO bounds. The agreement with the data is quantified by the $\chi^2$ computed with $1\sigma$ uncertainties. All points shown lie below the KamLAND-Zen limit and some points are above the future nEXO sensitivity in panel (h).}
    \label{fig:neutrino_oscillation_FDM}
\end{figure*}
The model also allows the lightest $Z_2$-odd particle to be the fermion $N_1$, which requires $D_{N1}<m_{\eta_I}$. The scans are otherwise performed as in the bosonic case, with the same parameters, the same $\chi^2$ criterion and the oscillation data again taken in normal ordering. Since the scan ranges are the same, $m_{\eta^\pm}>100$~GeV holds automatically here as well, and the only difference with respect to the bosonic case is the ordering $D_{N1}<m_{\eta_I}$.
The resulting neutrino phenomenology, shown in Fig.~\ref{fig:neutrino_oscillation_FDM}, is obtained under the condition $D_{N1}<m_{\eta_I}$ and has the same structure as in the bosonic case: the points reproduce the oscillation data in the planes of panels (a)--(d), populate a restricted region of the complex modulus in panel (e), and give predictions for the CP-violating phases in panels (f) and (g). In panel (e) the accepted points again cluster at isolated values of $\tau$. The clusters at $\mathrm{Im}\,\tau\simeq0.87$ are images of $\omega$ under shifts, the one at $\mathrm{Re}\,\tau\simeq0.5$ being also its mirror image $-\bar\omega$, while the lower clusters at $|\mathrm{Re}\,\tau|\simeq0.9$--$1.05$ are images of the large-$\mathrm{Im}\,\tau$ region. Both dark-matter scenarios therefore probe the same two physically distinct regions of the modulus. As shown in panel (h), all points that reproduce the oscillation data satisfy both the KamLAND-Zen limit and the cosmological bound from CMB data alone. Some points lie within the projected sensitivity of nEXO. On the other hand, some points are in tension with the stronger CMB+BAO bound.

\begin{figure*}[tbp]
    \centering
    \includegraphics[width=0.8\textwidth]{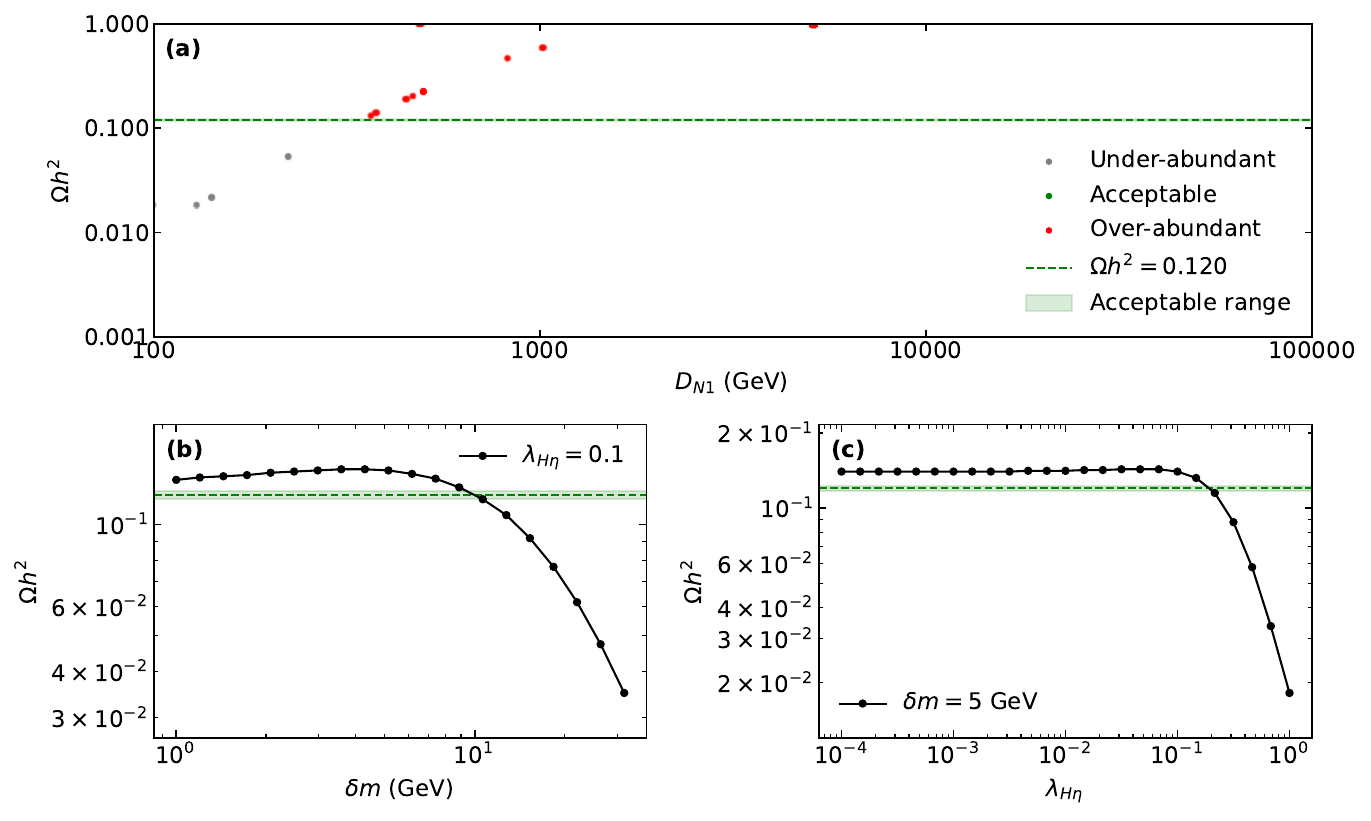}
    \caption{Relic density of $N_1$ for the fermionic dark-matter scenario. Panel (a) shows $\Omega h^2$ as a function of $D_{N1}$ for the points that reproduce the neutrino oscillation data and satisfy the lepton-flavour-violation bounds, with $\delta m=5$~GeV and $\lambda_{H\eta}=0.1$ held fixed. Panels (b) and (c) show $\Omega h^2$ for a representative point of panel (a), with one parameter varied while the other is kept at its benchmark value: (b) as a function of $\delta m=m_{\eta^\pm}-m_{\eta_I}$ at $\lambda_{H\eta}=0.1$, and (c) as a function of $\lambda_{H\eta}$ at $\delta m=5$~GeV. In all panels the green dashed line marks the observed value $\Omega h^2=0.120$ and the green band, labelled ``Acceptable'', the range $0.117\le\Omega h^2\le0.123$. In panel (a), underabundant points are shown in gray and overabundant points in red. The neutrino oscillation observables are unchanged along the curves in panels (b) and (c).}
    \label{fig:dm_FDM}
\end{figure*}

For the dark-matter phenomenology we follow the same procedure as for $\eta_I$: the model is implemented in \texttt{FeynRules}, exported in the \texttt{CalcHEP} format, and the relic density of $N_1$ is computed with \texttt{micrOMEGAs}. Fixing $\delta m$ and $\lambda_{H\eta}$ to the same benchmark values, $\delta m=5$~GeV and $\lambda_{H\eta}=0.1$, we obtain the relic density as a function of the dark-matter mass $D_{N1}$ shown in panel (a) of Fig.~\ref{fig:dm_FDM}, with the acceptable range $0.117\le\Omega h^2\le0.123$ as before. In contrast to the bosonic case, no point reproduces the observed relic density at the benchmark values, and all points are either overabundant or underabundant.
For the points we examined, $N_1$ is nearly degenerate with $\eta_I$ and $\eta_R$, so that the inert doublet remains in equilibrium with $N_1$ at freeze-out. The relic density is then governed mainly by the annihilation of the neutral inert scalars into $W^+W^-$ and $ZZ$ and by coannihilation involving $\eta^\pm$, while $N_1N_1$ annihilation through $\tilde y_\eta$ does not appear among the channels contributing more than $1\%$.

As in the bosonic case, the points can be brought into the acceptable range by tuning $\delta m$ and $\lambda_{H\eta}$, without altering the neutrino fit, since the oscillation observables are independent of both parameters. Panels (b) and (c) of Fig.~\ref{fig:dm_FDM} show, for a representative point, the relic density as a function of $\delta m$ at $\lambda_{H\eta}=0.1$ and as a function of $\lambda_{H\eta}$ at $\delta m=5$~GeV, respectively. Varying these two parameters, alone or together, brings the points into the acceptable range, and all the tuned points shown satisfy the lepton-flavour-violation bounds.

\section{CONCLUSION\label{sec:conclusion}}

We have studied a scotogenic model with non-holomorphic modular $A_4$ flavour symmetry, in which the Yukawa couplings are given by weight-zero polyharmonic Maa{\ss} forms and the flavour structure is controlled by a single complex modulus $\tau$, without flavon fields or supersymmetry. The modular weights forbid the operators $\overline{N_R}L_LH$ and $H\eta$, so that an accidental $Z_2$ symmetry forbids tree-level neutrino masses and guarantees the stability of the lightest $Z_2$-odd particle. Neutrino masses are generated at one loop, and the dark-matter candidate is either the fermion $N_1$ or the neutral inert scalar $\eta_I$. Using the latest NuFIT~v6.1 global fit, we confronted the model with the neutrino oscillation data in normal ordering, quantifying the agreement through a $\chi^2$ built from the three mixing angles and the two mass-squared differences with uncertainties at the $1\sigma$ level.

A central feature of our analysis is the parametrisation of the inert-scalar sector in terms of $m_{\eta_I}$, $\Delta m_\eta^2$, $\delta m$ and $\lambda_{H\eta}$. Only $m_{\eta_I}$ and $\Delta m_\eta^2$ enter the radiative neutrino mass, so that the oscillation fit fixes them for each point, while $\delta m$ and $\lambda_{H\eta}$ remain free and are determined only in the dark-matter analysis. We found that the points reproducing the oscillation data populate a restricted region of the complex modulus, which underlies the predictive power of the model in the lepton sector, and that all of them lie below the KamLAND-Zen limit while some of them are above future nEXO sensitivity for $m_{ee}$, and below the cosmological bounds on $\sum_i m_i$ from CMB data while some points are in tension with the CMB+BAO bound.

We then carried out an explicit study of the dark-matter phenomenology, with the relic density computed using \texttt{micrOMEGAs}, and imposed the LEP bound on the charged inert scalar together with the lepton-flavour-violation limits on $\ell_i\to\ell_j\gamma$. We found that, with these constraints taken into account, the model allows both bosonic and fermionic dark matter for neutrino oscillation data in normal ordering. For bosonic dark matter, a small fraction of the points reproduces the observed relic density at the benchmark values $\delta m=5$~GeV and $\lambda_{H\eta}=0.1$, and the remaining overabundant and underabundant points can be brought into the acceptable range by tuning $\delta m$, which controls the coannihilation with the charged scalar, and in some cases also $\lambda_{H\eta}$, which controls the Higgs-portal annihilation. For fermionic dark matter, no point reproduces the observed relic density at the benchmark values, but tuning the same two parameters brings the points into the acceptable range. Since $\delta m$ and $\lambda_{H\eta}$ do not enter the neutrino mass matrix, this tuning leaves the oscillation fit unchanged, and the tuned points continue to satisfy the lepton-flavour-violation bounds. The model therefore accommodates neutrino oscillation data, neutrinoless double-beta decay and cosmological bounds, lepton-flavour-violation limits and the observed dark-matter relic density within a single framework with a small number of free parameters.

\section*{Acknowledgments}
\noindent Priya and B. C. Chauhan would like to acknowledge the IUCAA for providing the HPC facility to carry out this work.

\appendix
\section{$A_4$ MULTIPLICATION RULES AND CONVENTIONS FOR MODULAR FORMS}
\label{sec:A4conv}

The irreducible representations of $A_4$ are $\mathbf{1},\mathbf{1}',\mathbf{1}'',\mathbf{3}$, with $\mathbf{1}'\otimes\mathbf{1}'=\mathbf{1}''$, $\mathbf{1}'\otimes\mathbf{1}''=\mathbf{1}$, $\mathbf{1}''\otimes\mathbf{1}''=\mathbf{1}'$. In the basis where
\begin{equation}
T=\mathrm{diag}(1,\omega,\omega^2),\qquad
S=\frac13\begin{pmatrix}-1&2&2\\2&-1&2\\2&2&-1\end{pmatrix},\qquad \omega=e^{2\pi i/3},
\end{equation}
the product of two triplets $\alpha=(\alpha_1,\alpha_2,\alpha_3)$ and $\beta=(\beta_1,\beta_2,\beta_3)$ decomposes as $\mathbf{3}\otimes\mathbf{3}=\mathbf{1}\oplus\mathbf{1}'\oplus\mathbf{1}''\oplus\mathbf{3}_S\oplus\mathbf{3}_A$ with
\begin{align}
(\alpha\beta)_{\mathbf 1}&=\alpha_1\beta_1+\alpha_2\beta_3+\alpha_3\beta_2,\nonumber\\
(\alpha\beta)_{\mathbf 1'}&=\alpha_3\beta_3+\alpha_1\beta_2+\alpha_2\beta_1,\nonumber\\
(\alpha\beta)_{\mathbf 1''}&=\alpha_2\beta_2+\alpha_1\beta_3+\alpha_3\beta_1,\nonumber\\
(\alpha\beta)_{\mathbf 3_S}&=\big(2\alpha_1\beta_1-\alpha_2\beta_3-\alpha_3\beta_2,\;2\alpha_3\beta_3-\alpha_1\beta_2-\alpha_2\beta_1,\;2\alpha_2\beta_2-\alpha_1\beta_3-\alpha_3\beta_1\big)^T,\nonumber\\
(\alpha\beta)_{\mathbf 3_A}&=\big(\alpha_2\beta_3-\alpha_3\beta_2,\;\alpha_1\beta_2-\alpha_2\beta_1,\;\alpha_3\beta_1-\alpha_1\beta_3\big)^T .
\end{align}

With $q=e^{2\pi i\tau}$ and $y=\mathrm{Im}\,\tau$, the components of the weight-zero triplet $Y_3^{(0)}$ have the expansions
\begin{align}
Y^{(0)}_{3,1}&=y-\frac{3e^{-4\pi y}}{\pi q}-\frac{9e^{-8\pi y}}{2\pi q^2}-\frac{e^{-12\pi y}}{\pi q^3}-\frac{21e^{-16\pi y}}{4\pi q^4}+\cdots\nonumber\\
&\quad-\frac{9\log3}{4\pi}-\frac{3q}{\pi}-\frac{9q^2}{2\pi}-\frac{q^3}{\pi}-\frac{21q^4}{4\pi}+\cdots,\\[2mm]
Y^{(0)}_{3,2}&=\frac{27q^{1/3}e^{\pi y/3}}{\pi}\Big(\frac{e^{-3\pi y}}{4q}+\frac{e^{-7\pi y}}{5q^2}+\frac{5e^{-11\pi y}}{16q^3}+\cdots\Big)\nonumber\\
&\quad+\frac{9q^{1/3}}{2\pi}\Big(1+\frac{7q}{4}+\frac{8q^2}{7}+\frac{9q^3}{5}+\cdots\Big),\\[2mm]
Y^{(0)}_{3,3}&=\frac{9q^{2/3}e^{2\pi y/3}}{2\pi}\Big(\frac{e^{-2\pi y}}{q}+\frac{7e^{-6\pi y}}{4q^2}+\frac{8e^{-10\pi y}}{7q^3}+\cdots\Big)\nonumber\\
&\quad+\frac{27q^{2/3}}{\pi}\Big(\frac14+\frac{q}{5}+\frac{5q^2}{16}+\frac{2q^3}{11}+\cdots\Big).
\end{align}
\section{EQUIVALENT VALUES OF THE MODULUS}
\label{sec:tauequiv}

The modulus $\tau$ labels the theory, but several values of $\tau$ describe the same physics. The simplest equivalence is the shift $T:\tau\to\tau+1$. From the expansions of Appendix~\ref{sec:A4conv} one finds $Y_3^{(0)}(\tau+1)=\rho(T)\,Y_3^{(0)}(\tau)$ with $\rho(T)=\mathrm{diag}(1,\omega,\omega^2)$, which we also verified numerically. Since $L_L$, $e_R$ and $N_R$ are all triplets, this phase matrix is absorbed by a redefinition of the lepton fields, so that for the same couplings the charged-lepton and neutrino mass matrices at $\tau+1$ give the same masses, mixing angles and phases as at $\tau$. Because $T^3=\mathbb{I}$, $Y_3^{(0)}(\tau+3)=Y_3^{(0)}(\tau)$, so the interval $-1.5\le\mathrm{Re}\,\tau\le1.5$ of our scans covers exactly one period and $\mathrm{Re}\,\tau=-1.5$ and $+1.5$ are the same line.

The inversion $S:\tau\to-1/\tau$ is also a symmetry. All Yukawa and Majorana mass terms of Eq.~\eqref{eq:yukawa} have total modular weight zero, and the weight-zero triplet satisfies $Y_3^{(0)}(-1/\tau)=\rho(S)\,Y_3^{(0)}(\tau)$, which we again verified numerically with the expansions above. The fields transform with the unitary matrix $\rho(S)$ and with weight factors that do not affect the physical observables, so the Yukawa couplings $y_\ell$, $y_{\ell3}$, $y'_{\ell3}$, $y_D$, $y_{D3}$, $y'_{D3}$ and the Majorana parameters take the same values at equivalent points. The scalar-potential coupling $\lambda_0$ multiplies a weight $-2$ form, so its value differs between equivalent points; since it is scanned freely, this does not change the set of accessible observables. Together, $S$ and $T$ generate $SL(2,\mathbb{Z})$, so every $\tau$ is equivalent to a point of the fundamental domain $|\mathrm{Re}\,\tau|\le\tfrac12$, $|\tau|\ge1$. In addition, the expansions give $Y_3^{(0)}(-\bar\tau)=Y_3^{(0)}(\tau)^*$, so that for real couplings, as in our scans, the map $\tau\to-\bar\tau$ complex-conjugates the mass matrices: the masses and mixing angles are unchanged, while $\delta_{CP}$, $\alpha_{21}$ and $\alpha_{31}$ change sign.

Two points of the fundamental domain are special. The point $\omega=-\tfrac12+i\tfrac{\sqrt3}{2}$ is fixed by $ST$, since $-1/(\omega+1)=\omega$, and therefore preserves a residual $Z_3$ symmetry; the point $\omega+1=-\bar\omega$ is equivalent to $\omega$ both by $S$ and by the mirror map. At large $\mathrm{Im}\,\tau$ the triplet approaches the direction $(1,0,0)$, which is invariant under $T$, with corrections of a few percent; for example $Y_3^{(0)}\simeq(0.96,\,0.04,\,0.04)$ at $\tau=1.75\,i$. The flavour structure is therefore approximately constrained by a residual symmetry in this region as well.

Since our scan box contains several copies of the fundamental domain, the accepted points fall into two physically distinct classes. The neighbourhood of $\omega$ appears at $\mathrm{Re}\,\tau\simeq-1.5,\,-0.5,\,0.5$ with $\mathrm{Im}\,\tau\simeq0.87$, where the images at $\pm1.5$ coincide. The region of large $\mathrm{Im}\,\tau$ is also seen at $\mathrm{Im}\,\tau\simeq0.5$--$0.6$ through $\tau\to-1/(\tau\pm1)$; for instance $\tau=-0.909+0.563\,i$ is mapped to $-0.28+1.73\,i$. All results in the main text are therefore understood modulo these equivalences.


\bibliographystyle{unsrt}
\bibliography{reference.bib} 

@article{Cheshta:2026fls,
    author = "Cheshta and Priya and Dutt, Suneel and Chauhan, B. C.",
    title = "{A Type-I seesaw framework with non-holomorphic modular symmetry}",
    eprint = "2604.18070",
    archivePrefix = "arXiv",
    primaryClass = "hep-ph",
    doi = "10.1140/epjc/s10052-026-15675-3",
    journal = "Eur. Phys. J. C",
    volume = "86",
    number = "4",
    pages = "400",
    year = "2026"
}

@article{Loualidi:2026pld,
    author = "Loualidi, Mohamed Amin and Miskaoui, Mohamed and Nasri, Salah",
    title = "{Radiative Neutrino Mass in a Nonholomorphic $T'$ Modular Invariant Model}",
    eprint = "2606.11346",
    archivePrefix = "arXiv",
    primaryClass = "hep-ph",
    month = "6",
    year = "2026"
}

@article{Abbas:2026siw,
    author = "Abbas, Mohammed",
    title = "{Lepton masses and mixing in non-holomorphic modular $A_4$ with universal couplings}",
    eprint = "2604.16130",
    archivePrefix = "arXiv",
    primaryClass = "hep-ph",
    month = "4",
    year = "2026"
}

@article{CarcamoHernandez:2026pis,
    author = "C{\'a}rcamo Hern{\'a}ndez, A. E. and Echeverria-Puentes, J. and K. N., Vishnudath and Kovalenko, Sergey and Salinas-Arizmendi, Daniel and Vaquera-Araujo, Carlos A.",
    title = "{Two-loop neutrino mass model with modular $S_4$ symmetry}",
    eprint = "2605.11097",
    archivePrefix = "arXiv",
    primaryClass = "hep-ph",
    month = "5",
    year = "2026"
}

@article{deMedeirosVarzielas:2026lfw,
    author = "de Medeiros Varzielas, Ivo and Paiva, Manuel",
    title = {{Quark masses and mixing from Modular $S'_4$ with Canonical K{\"a}hler Effects}},
    eprint = "2604.01422",
    archivePrefix = "arXiv",
    primaryClass = "hep-ph",
    month = "4",
    year = "2026"
}

@article{deMedeirosVarzielas:2025byb,
    author = "de Medeiros Varzielas, Ivo and Liu, Ming-Shau and Sengupta, Amartya and Talbert, Jim",
    title = "{Residual Symmetries and Scalar Multiplet Vacuum Alignment in Non-Abelian Flavour Models}",
    eprint = "2512.19789",
    archivePrefix = "arXiv",
    primaryClass = "hep-ph",
    reportNumber = "LA-UR-25-31854",
    month = "12",
    year = "2025"
}

@article{Ishiguro:2022pde,
    author = "Ishiguro, Keiya and Okada, Hiroshi and Otsuka, Hajime",
    title = "{Residual flavor symmetry breaking in the landscape of modular flavor models}",
    eprint = "2206.04313",
    archivePrefix = "arXiv",
    primaryClass = "hep-ph",
    reportNumber = "KEK-TH-2430, APCTP Pre2022 - 011, KYUSHU-HET-243",
    doi = "10.1007/JHEP09(2022)072",
    journal = "JHEP",
    volume = "09",
    pages = "072",
    year = "2022"
}

@article{Behera:2026tmo,
    author = "Behera, Mitesh Kumar and Das, Jaydeb and Mondal, Niloy",
    title = "{A Non-Holomorphic Modular $A_4$ Framework for Resonant Leptogenesis with Gravitational Wave Signatures}",
    eprint = "2607.18803",
    archivePrefix = "arXiv",
    primaryClass = "hep-ph",
    month = "7",
    year = "2026"
}

@article{Li:2026ory,
    author = "Li, Cai-Chang and Lu, Jun-Nan and Gao, Xiang-Yan and Weng, Ming-Hua and Ding, Gui-Jun",
    title = "{Non-holomorphic $S^{\prime}_{4}$ modular symmetry for leptons and leptogenesis}",
    eprint = "2606.24277",
    archivePrefix = "arXiv",
    primaryClass = "hep-ph",
    month = "6",
    year = "2026"
}

@article{CarcamoHernandez:2025atu,
    author = "C{\'a}rcamo Hern{\'a}ndez, A. E. and de Medeiros Varzielas, Ivo and King, S. F. and K. N., Vishnudath",
    title = "{Modular S3 flavored Pati-Salam model with a two family seesaw mechanism}",
    eprint = "2503.14610",
    archivePrefix = "arXiv",
    primaryClass = "hep-ph",
    doi = "10.1103/6yjc-3wlc",
    journal = "Phys. Rev. D",
    volume = "112",
    number = "9",
    pages = "095042",
    year = "2025"
}

@article{Kobayashi:2021pav,
    author = "Kobayashi, Tatsuo and Otsuka, Hajime and Tanimoto, Morimitsu and Yamamoto, Kei",
    title = "{Modular symmetry in the SMEFT}",
    eprint = "2112.00493",
    archivePrefix = "arXiv",
    primaryClass = "hep-ph",
    reportNumber = "EPHOU-21-014, CTPU-PTC-21-39, KEK-TH-2369, HUPD-2113",
    doi = "10.1103/PhysRevD.105.055022",
    journal = "Phys. Rev. D",
    volume = "105",
    number = "5",
    pages = "055022",
    year = "2022"
}

@article{Zhang:2026kyy,
    author = "Zhang, Xianshuo and Reyimuaji, Yakefu",
    title = "{Neutrino Mass and Leptogenesis in the Non-SUSY Modular $A^\prime_5$ Inverse Seesaw}",
    eprint = "2603.19104",
    archivePrefix = "arXiv",
    primaryClass = "hep-ph",
    month = "3",
    year = "2026"
}

@article{Majhi:2026jdk,
    author = "Majhi, Rudra and Behera, Mitesh Kumar and Mohanta, Rukmani",
    title = "{A Predictive Non-Holomorphic Modular $A_4$ Linear Seesaw Framework Testable at DUNE}",
    eprint = "2602.23018",
    archivePrefix = "arXiv",
    primaryClass = "hep-ph",
    month = "2",
    year = "2026"
}

@article{Mishra:2023ekx,
    author = "Mishra, Priya and Behera, Mitesh Kumar and Panda, Papia and Ghosh, Monojit and Mohanta, Rukmani",
    title = "{Exploring models with modular symmetry in neutrino oscillation experiments}",
    eprint = "2305.08576",
    archivePrefix = "arXiv",
    primaryClass = "hep-ph",
    doi = "10.1007/JHEP09(2023)144",
    journal = "JHEP",
    volume = "09",
    pages = "144",
    year = "2023"
}

@article{Kumar:2023moh,
    author = "Kumar, Ranjeet and Mishra, Priya and Behera, Mitesh Kumar and Mohanta, Rukmani and Srivastava, Rahul",
    title = "{Predictions from scoto-seesaw with A4 modular symmetry}",
    eprint = "2310.02363",
    archivePrefix = "arXiv",
    primaryClass = "hep-ph",
    doi = "10.1016/j.physletb.2024.138635",
    journal = "Phys. Lett. B",
    volume = "853",
    pages = "138635",
    year = "2024"
}

@article{NuFit61,
    title = "{v6.1: Three-neutrino fit based on data available in November 2025}",
    doi = "http://www.nu-fit.org/?q=node/309",
    journal = "NuFit-6.1",
    year = "2025"
}

@article{Feruglio:2017spp,
    author = "Feruglio, Ferruccio",
    editor = "Levy, Aharon and Forte, Stefano and Ridolfi, Giovanni",
    title = "{Are neutrino masses modular forms?}",
    booktitle = "{From My Vast Repertoire ...}: {Guido Altarelli's Legacy}",
    eprint = "1706.08749",
    archivePrefix = "arXiv",
    primaryClass = "hep-ph",
    reportNumber = "DFPD-2017-TH-09",
    doi = "10.1142/9789813238053_0012",
    pages = "227--266",
    year = "2019"
}

@article{Tapender:2026ets,
    author = "Tapender and Verma, Surender",
    title = "{Tri-Resonant Leptogenesis in a Non-Holomorphic Modular A$_4$ Scotogenic Model}",
    eprint = "2602.17243",
    archivePrefix = "arXiv",
    primaryClass = "hep-ph",
    month = "2",
    year = "2026"
}

@article{Tapender:2025cfc,
    author = "Tapender and Kumar, Sanjeev and Verma, Surender",
    title = "{Unveiling the Impact of Generalized CP Symmetry on 2HDM and Neutrino Physics}",
    doi = "10.1007/978-981-97-7441-8_26",
    journal = "Springer Proc. Phys.",
    volume = "361",
    pages = "261--272",
    year = "2025"
}

@article{Tapender:2023kdk,
    author = "Tapender and Kumar, Sanjeev and Verma, Surender",
    title = "{Neutrino phenomenology in a model with generalized CP symmetry within type-I seesaw framework}",
    eprint = "2309.04242",
    archivePrefix = "arXiv",
    primaryClass = "hep-ph",
    doi = "10.1103/PhysRevD.109.015004",
    journal = "Phys. Rev. D",
    volume = "109",
    number = "1",
    pages = "015004",
    year = "2024"
}

@article{Nasri:2026nbf,
    author = "Nasri, Salah and Singh, Labh and Tapender and Verma, Surender",
    title = "{Dark-Portal Leptogenesis in a Non-Holomorphic Modular Scoto-Seesaw Model}",
    eprint = "2601.06435",
    archivePrefix = "arXiv",
    primaryClass = "hep-ph",
    month = "1",
    year = "2026"
}

@article{Ohki:2020bpo,
    author = "Ohki, Hiroshi and Uemura, Shohei and Watanabe, Risa",
    title = "{Modular flavor symmetry on a magnetized torus}",
    eprint = "2003.04174",
    archivePrefix = "arXiv",
    primaryClass = "hep-th",
    doi = "10.1103/PhysRevD.102.085008",
    journal = "Phys. Rev. D",
    volume = "102",
    number = "8",
    pages = "085008",
    year = "2020"
}

@article{deAdelhartToorop:2011re,
    author = "de Adelhart Toorop, Reinier and Feruglio, Ferruccio and Hagedorn, Claudia",
    title = "{Finite Modular Groups and Lepton Mixing}",
    eprint = "1112.1340",
    archivePrefix = "arXiv",
    primaryClass = "hep-ph",
    reportNumber = "DFPD-11-TH-18",
    doi = "10.1016/j.nuclphysb.2012.01.017",
    journal = "Nucl. Phys. B",
    volume = "858",
    pages = "437--467",
    year = "2012"
}

@article{Super-Kamiokande:1998kpq,
    author = "Fukuda, Y. and others",
    collaboration = "Super-Kamiokande",
    title = "{Evidence for oscillation of atmospheric neutrinos}",
    eprint = "hep-ex/9807003",
    archivePrefix = "arXiv",
    reportNumber = "BU-98-17, ICRR-REPORT-422-98-18, UCI-98-8, KEK-PREPRINT-98-95, LSU-HEPA-5-98, UMD-98-003, SBHEP-98-5, TKU-PAP-98-06, TIT-HPE-98-09",
    doi = "10.1103/PhysRevLett.81.1562",
    journal = "Phys. Rev. Lett.",
    volume = "81",
    pages = "1562--1567",
    year = "1998"
}

@article{Priya:2025wdm,
    author = "Priya and Singh, Labh and Chauhan, B. C. and Verma, Surender",
    title = "{Type-III seesaw in non-holomorphic modular symmetry and leptogenesis}",
    eprint = "2508.05047",
    archivePrefix = "arXiv",
    primaryClass = "hep-ph",
    doi = "10.1007/JHEP01(2026)036",
    journal = "JHEP",
    volume = "01",
    pages = "036",
    year = "2026"
}

@article{SNO:2002tuh,
    author = "Ahmad, Q. R. and others",
    collaboration = "SNO",
    title = "{Direct evidence for neutrino flavor transformation from neutral current interactions in the Sudbury Neutrino Observatory}",
    eprint = "nucl-ex/0204008",
    archivePrefix = "arXiv",
    doi = "10.1103/PhysRevLett.89.011301",
    journal = "Phys. Rev. Lett.",
    volume = "89",
    pages = "011301",
    year = "2002"
}

@article{Priya:2026ytf,
    author = "Priya and Arora, Simran and Chauhan, B. C.",
    title = "{Generalized CP Symmetry and Texture Zero in Trimaximal Mixing Matrix}",
    doi = "10.1007/978-981-96-4986-0_64",
    journal = "Springer Proc. Phys.",
    volume = "322",
    pages = "397--400",
    year = "2026"
}

@article{Grossman:1998jj,
    author = "Grossman, Yuval and Nir, Yosef and Shadmi, Yael",
    title = "{Large mixing and large hierarchy between neutrinos with Abelian flavor symmetries}",
    eprint = "hep-ph/9808355",
    archivePrefix = "arXiv",
    reportNumber = "SLAC-PUB-7911, WIS-98-21-DPP",
    doi = "10.1088/1126-6708/1998/10/007",
    journal = "JHEP",
    volume = "10",
    pages = "007",
    year = "1998"
}

@article{Barman:2025bru,
    author = "Barman, Animesh",
    title = "{Phenomenology of Neutrino Masses and Mixing with Discrete Flavour Symmetry in the context of the latest neutrino oscillation data}",
    school = "Tezpur University, Department of Physics, India",
    year = "2025"
}

@article{Chauhan:2023faf,
    author = "Chauhan, Garv and Dev, P. S. Bhupal and Dubovyk, Ievgen and Dziewit, Bartosz and Flieger, Wojciech and Grzanka, Krzysztof and Gluza, Janusz and Karmakar, Biswajit and Zi{\k{e}}ba, Szymon",
    title = "{Phenomenology of lepton masses and mixing with discrete flavor symmetries}",
    eprint = "2310.20681",
    archivePrefix = "arXiv",
    primaryClass = "hep-ph",
    doi = "10.1016/j.ppnp.2024.104126",
    journal = "Prog. Part. Nucl. Phys.",
    volume = "138",
    pages = "104126",
    year = "2024"
}

@article{Priya:2025khf,
    author = "Priya and Arora, Simran and Chauhan, B. C.",
    title = "{On the Generalized CP Symmetry, One Zero Texture in Neutrino Mass Matrix and Neutrinoless Double Beta Decay}",
    eprint = "2501.00776",
    archivePrefix = "arXiv",
    primaryClass = "hep-ph",
    month = "1",
    year = "2025"
}

@article{Singh:2024imk,
    author = "Singh, Labh and Kashav, Monal and Verma, Surender",
    title = "{Minimal type-I Dirac seesaw and leptogenesis under A4 modular invariance}",
    eprint = "2405.07165",
    archivePrefix = "arXiv",
    primaryClass = "hep-ph",
    doi = "10.1016/j.nuclphysb.2024.116666",
    journal = "Nucl. Phys. B",
    volume = "1007",
    pages = "116666",
    year = "2024"
}

@article{Kashav:2021zir,
    author = "Kashav, Monal and Verma, Surender",
    title = "{Broken scaling neutrino mass matrix and leptogenesis based on A$_{4}$ modular invariance}",
    eprint = "2103.07207",
    archivePrefix = "arXiv",
    primaryClass = "hep-ph",
    doi = "10.1007/JHEP09(2021)100",
    journal = "JHEP",
    volume = "09",
    pages = "100",
    year = "2021"
}

@article{Kashav:2024lkr,
    author = "Kashav, Monal and Patel, Ketan M.",
    title = "{Residual flavor symmetries at the modular self-dual point and constraints on neutrino masses and mixing}",
    eprint = "2410.00565",
    archivePrefix = "arXiv",
    primaryClass = "hep-ph",
    doi = "10.1103/PhysRevD.111.015010",
    journal = "Phys. Rev. D",
    volume = "111",
    number = "1",
    pages = "015010",
    year = "2025"
}

@article{Feruglio:2021dte,
    author = "Feruglio, Ferruccio and Gherardi, Valerio and Romanino, Andrea and Titov, Arsenii",
    title = "{Modular invariant dynamics and fermion mass hierarchies around $\tau = i$}",
    eprint = "2101.08718",
    archivePrefix = "arXiv",
    primaryClass = "hep-ph",
    reportNumber = "SISSA 05/2021/FISI, FTUV-21-0119.3119, IFIC/21-02",
    doi = "10.1007/JHEP05(2021)242",
    journal = "JHEP",
    volume = "05",
    pages = "242",
    year = "2021"
}

@article{Nanda:2025lem,
    author = "Nanda, Swaraj Kumar and Ricky Devi, Maibam and Patra, Sudhanwa",
    title = "{Non-Holomorphic $A_4$ Modular Symmetry in Type-I Seesaw: Implications for Neutrino Masses and Leptogenesis}",
    eprint = "2509.22108",
    archivePrefix = "arXiv",
    primaryClass = "hep-ph",
    month = "9",
    year = "2025"
}

@article{Nomura:2025raf,
    author = "Nomura, Takaaki and Okada, Hiroshi",
    title = "{Neutrino mass model at a three-loop level from a non-holomorphic modular $A_4$ symmetry}",
    eprint = "2506.02639",
    archivePrefix = "arXiv",
    primaryClass = "hep-ph",
    month = "6",
    year = "2025"
}

@article{Nomura:2025ovm,
    author = "Nomura, Takaaki and Okada, Hiroshi and Wang, Xing-Yu",
    title = "{A radiative neutrino mass model with leptoquarks under non-holomorphic modular A$_{4}$ symmetry}",
    eprint = "2504.21404",
    archivePrefix = "arXiv",
    primaryClass = "hep-ph",
    doi = "10.1007/JHEP09(2025)163",
    journal = "JHEP",
    volume = "09",
    pages = "163",
    year = "2025"
}

@article{Nomura:2024vzw,
    author = "Nomura, Takaaki and Okada, Hiroshi and Popov, Oleg",
    title = "{Non-holomorphic modular A4 symmetric scotogenic model}",
    eprint = "2409.12547",
    archivePrefix = "arXiv",
    primaryClass = "hep-ph",
    doi = "10.1016/j.physletb.2024.139171",
    journal = "Phys. Lett. B",
    volume = "860",
    pages = "139171",
    year = "2025"
}

@article{Nomura:2024atp,
    author = "Nomura, Takaaki and Okada, Hiroshi",
    title = "{Type-II seesaw of a non-holomorphic modular A4 symmetry}",
    eprint = "2408.01143",
    archivePrefix = "arXiv",
    primaryClass = "hep-ph",
    doi = "10.1016/j.physletb.2025.139763",
    journal = "Phys. Lett. B",
    volume = "868",
    pages = "139763",
    year = "2025"
}

@article{Qu:2024rns,
    author = "Qu, Bu-Yao and Ding, Gui-Jun",
    title = "{Non-holomorphic modular flavor symmetry}",
    eprint = "2406.02527",
    archivePrefix = "arXiv",
    primaryClass = "hep-ph",
    doi = "10.1007/JHEP08(2024)136",
    journal = "JHEP",
    volume = "08",
    pages = "136",
    year = "2024"
}

@article{Kumar:2024uxn,
    author = "Kumar, Bhabana and Das, Mrinal Kumar",
    title = "{Study of neutrino phenomenology and 0{\ensuremath{\nu}}{\ensuremath{\beta}}{\ensuremath{\beta}} decay using polyharmonic Maa{\ensuremath{\beta}} forms}",
    eprint = "2405.10586",
    archivePrefix = "arXiv",
    primaryClass = "hep-ph",
    doi = "10.1142/S0217751X25500903",
    journal = "Int. J. Mod. Phys. A",
    volume = "40",
    number = "23",
    pages = "2550090",
    year = "2025"
}

@article{Priya:2026ehe,
    author = "Priya and Chauhan, B. C. and Kumar, Deepak and Nomura, Takaaki",
    title = "{Predictions of Modular Symmetry Fixed Points on Neutrino Masses, Mixing, and Leptogenesis}",
    eprint = "2604.04585",
    archivePrefix = "arXiv",
    primaryClass = "hep-ph",
    month = "4",
    year = "2026"
}

@article{Ma:2006km,
    author = "Ma, Ernest",
    title = "{Verifiable radiative seesaw mechanism of neutrino mass and dark matter}",
    eprint = "hep-ph/0601225",
    archivePrefix = "arXiv",
    reportNumber = "UCRHEP-T403",
    doi = "10.1103/PhysRevD.73.077301",
    journal = "Phys. Rev. D",
    volume = "73",
    pages = "077301",
    year = "2006"
}

@article{Kumar:2025nut,
    author = "Kumar, Bhabana and Das, Mrinal Kumar",
    title = "{Neutrino phenomenology and Dark matter in a left-right asymmetric model with non-holomorphic modular $A_{4}$ group}",
    eprint = "2509.01205",
    archivePrefix = "arXiv",
    primaryClass = "hep-ph",
    month = "9",
    year = "2025"
}

@article{Dey:2025zld,
    author = "Dey, Manash",
    title = "{The Seesaw Evaded Modular Dirac Framework}",
    eprint = "2509.10373",
    archivePrefix = "arXiv",
    primaryClass = "hep-ph",
    month = "9",
    year = "2025"
}

@article{Gao:2025jlw,
    author = "Gao, Xiang-Yan and Li, Cai-Chang",
    title = "{Minimal lepton models with non-holomorphic modular $A_{4}$ symmetry}",
    eprint = "2512.07158",
    archivePrefix = "arXiv",
    primaryClass = "hep-ph",
    month = "12",
    year = "2025"
}

@article{Jangid:2025thp,
    author = "Jangid, Shilpa and Okada, Hiroshi",
    title = "{A radiative seesaw model in a non-invertible selection rule with the assistance of a non-holomorphic modular $A_4$ symmetry}",
    eprint = "2510.17292",
    archivePrefix = "arXiv",
    primaryClass = "hep-ph",
    month = "10",
    year = "2025"
}

@article{Li:2025kcr,
    author = "Li, Cai-Chang and Ding, Gui-Jun",
    title = "{Lepton models from non-holomorphic $A^{\prime}_{5}$ modular flavor symmetry}",
    eprint = "2509.15183",
    archivePrefix = "arXiv",
    primaryClass = "hep-ph",
    month = "9",
    year = "2025"
}

@article{Zhang:2025dsa,
    author = "Zhang, Xianshuo and Reyimuaji, Yakefu",
    title = "{Inverse seesaw model in nonholomorphic modular A4 flavor symmetry}",
    eprint = "2507.06945",
    archivePrefix = "arXiv",
    primaryClass = "hep-ph",
    doi = "10.1103/17p3-bw5r",
    journal = "Phys. Rev. D",
    volume = "112",
    number = "7",
    pages = "075050",
    year = "2025"
}

@article{Abbas:2025nlv,
    author = "Abbas, Mohammed",
    title = "{Lepton Masses and Mixing in Nonholomorphic Modular A4 Symmetry}",
    doi = "10.31526/PHEP.2025.07",
    journal = "PHEP",
    volume = "2025",
    pages = "7",
    year = "2025"
}

@article{Loualidi:2025tgw,
    author = "Loualidi, Mohamed Amin and Miskaoui, Mohamed and Nasri, Salah",
    title = "{Nonholomorphic A4 modular invariance for fermion masses and mixing in SU(5) GUT}",
    eprint = "2503.12594",
    archivePrefix = "arXiv",
    primaryClass = "hep-ph",
    doi = "10.1103/1py2-cmfx",
    journal = "Phys. Rev. D",
    volume = "112",
    number = "1",
    pages = "015008",
    year = "2025"
}

@article{Okada:2025jjo,
    author = "Okada, Hiroshi and Orikasa, Yuta",
    title = "{A radiative seesaw in a non-holomorphic modular $S_3$ flavor symmetry}",
    eprint = "2501.15748",
    archivePrefix = "arXiv",
    primaryClass = "hep-ph",
    month = "1",
    year = "2025"
}

@article{Nomura:2024nwh,
    author = "Nomura, Takaaki and Okada, Hiroshi",
    title = "{Zee model in a non-holomorphic modular A4 symmetry}",
    eprint = "2412.18095",
    archivePrefix = "arXiv",
    primaryClass = "hep-ph",
    doi = "10.1016/j.physletb.2025.139618",
    journal = "Phys. Lett. B",
    volume = "867",
    pages = "139618",
    year = "2025"
}

@article{Li:2024svh,
    author = "Li, Cai-Chang and Lu, Jun-Nan and Ding, Gui-Jun",
    title = "{Non-holomorphic modular A$_{5}$ symmetry for lepton masses and mixing}",
    eprint = "2410.24103",
    archivePrefix = "arXiv",
    primaryClass = "hep-ph",
    doi = "10.1007/JHEP12(2024)189",
    journal = "JHEP",
    volume = "12",
    pages = "189",
    year = "2024"
}

@article{Ding:2024inn,
    author = "Ding, Gui-Jun and Lu, Jun-Nan and Petcov, S. T. and Qu, Bu-Yao",
    title = "{Non-holomorphic modular S$_{4}$ lepton flavour models}",
    eprint = "2408.15988",
    archivePrefix = "arXiv",
    primaryClass = "hep-ph",
    reportNumber = "SISSA 17/2024/FISI",
    doi = "10.1007/JHEP01(2025)191",
    journal = "JHEP",
    volume = "01",
    pages = "191",
    year = "2025"
}

@article{nEXO:2021ujk,
    author = "Adhikari, G. and others",
    collaboration = "nEXO",
    title = "{nEXO: neutrinoless double beta decay search beyond 10$^{28}$ year half-life sensitivity}",
    eprint = "2106.16243",
    archivePrefix = "arXiv",
    primaryClass = "nucl-ex",
    doi = "10.1088/1361-6471/ac3631",
    journal = "J. Phys. G",
    volume = "49",
    number = "1",
    pages = "015104",
    year = "2022"
}

@article{Planck:2018vyg,
    author = "Aghanim, N. and others",
    collaboration = "Planck",
    title = "{Planck 2018 results. VI. Cosmological parameters}",
    eprint = "1807.06209",
    archivePrefix = "arXiv",
    primaryClass = "astro-ph.CO",
    doi = "10.1051/0004-6361/201833910",
    journal = "Astron. Astrophys.",
    volume = "641",
    pages = "A6",
    year = "2020",
    note = "[Erratum: Astron.Astrophys. 652, C4 (2021)]"
}

@article{KamLAND:2002uet,
    author = "Eguchi, K. and others",
    collaboration = "KamLAND",
    title = "{First results from KamLAND: Evidence for reactor anti-neutrino disappearance}",
    eprint = "hep-ex/0212021",
    archivePrefix = "arXiv",
    doi = "10.1103/PhysRevLett.90.021802",
    journal = "Phys. Rev. Lett.",
    volume = "90",
    pages = "021802",
    year = "2003"
}

@article{KamLAND-Zen:2024eml,
    author = "Abe, S. and others",
    collaboration = "KamLAND-Zen",
    title = "{Search for Majorana Neutrinos with the Complete KamLAND-Zen Dataset}",
    eprint = "2406.11438",
    archivePrefix = "arXiv",
    primaryClass = "hep-ex",
    month = "6",
    year = "2024"
}

@article{toma2014lepton,
  title={Lepton flavor violation in the scotogenic model},
  author={Toma, Takashi and Vicente, Avelino},
  journal={Journal of High Energy Physics},
  volume={2014},
  number={1},
  pages={1--27},
  year={2014},
  publisher={Springer}
}

@article{alloul2014feynrules,
  title={FeynRules 2.0—A complete toolbox for tree-level phenomenology},
  author={Alloul, Adam and Christensen, Neil D and Degrande, C{\'e}line and Duhr, Claude and Fuks, Benjamin},
  journal={Computer Physics Communications},
  volume={185},
  number={8},
  pages={2250--2300},
  year={2014},
  publisher={Elsevier}
}

@article{belanger2026micromegas,
  title={micrOMEGAs 7: Beyond standard cosmology},
  author={Belanger, G and Belyaev, A and Bernal, N and Boudjema, F and Chakraborti, S and Goudelis, A and Pukhov, A},
  journal={arXiv preprint arXiv:2606.06645},
  year={2026}
}

@article{nomura2019modular,
  title={A modular A4 symmetric model of dark matter and neutrino},
  author={Nomura, Takaaki and Okada, Hiroshi},
  journal={Physics Letters B},
  volume={797},
  pages={134799},
  year={2019},
  publisher={Elsevier}
}

@article{aghanim2020planck,
  title={Planck 2018 results},
  author={Aghanim, Nabila and Akrami, Yashar and Arroja, F and Ashdown, M and Aumont, J and Baccigalupi, C and Ballardini, M and Banday, AJ and Barreiro, RB and Bartolo, N and others},
  year={2020}
}

@article{adame2025desi,
  title={DESI 2024 VI: cosmological constraints from the measurements of baryon acoustic oscillations},
  author={Adame, Armand G and Aguilar, J and Ahlen, Steven and Alam, Shadab and Alexander, David M and Alvarez, M and Alves, Ot{\'a}vio and Anand, Abhijeet and Andrade, Uendert and Armengaud, Eric and others},
  journal={Journal of Cosmology and Astroparticle Physics},
  volume={2025},
  number={02},
  pages={021},
  year={2025},
  publisher={IOP Publishing}
}
\end{document}